\documentclass[iop]{emulateapj}

\usepackage{braket}

\shorttitle{Planet Traps and the Formation of Super-Earths}
\shortauthors{Hasegawa}

\begin{document}

\title{Super-Earths as Failed Cores in Orbital Migration Traps}

\author{Yasuhiro Hasegawa\altaffilmark{1,2,3,4}}
\affil{$^1$Institute of Astronomy and Astrophysics, Academia Sinica (ASIAA), Taipei 10641, Taiwan}
\affil{$^2$Division of Theoretical Astronomy, National Astronomical Observatory of Japan, Osawa, Mitaka, Tokyo 181-8588, Japan}
\affil{$^3$Jet Propulsion Laboratory, California Institute of Technology, Pasadena, CA 91109, USA}
\email{yasuhiro@caltech.edu}

\altaffiltext{4}{EACOA fellow}

\begin{abstract}
We explore whether close-in super-Earths were formed 
as rocky bodies that failed to grow fast enough to become the cores of gas giants before the natal protostellar disk dispersed.  
We model the failed cores' inward orbital migration in the low-mass or type I regime, 
to stopping points at distances where the tidal interaction with the protostellar disk applies zero net torque.  
The three kinds of migration traps considered are those due to the dead zone's outer edge, the ice line, 
and the transition from accretion to starlight as the disk's main heat source.  
As the disk disperses, the traps move toward final positions near or just outside 1~au.  
Planets at this location exceeding about 3~M$_\oplus$ open a gap, decouple from their host trap, 
and migrate inward in the high-mass or type II regime to reach the vicinity of the star.  
We synthesize the population of planets formed in this scenario, 
finding that some fraction of the observed super-Earths can be failed cores.  
Most super-Earths formed this way have more than 4~M$_\oplus$, 
so their orbits when the disk disperses are governed by type II migration.  
These planets have solid cores surrounded by gaseous envelopes.  
Their subsequent photoevaporative mass loss is most effective for masses originally below about 6 M$_\oplus$.  
The failed core scenario suggests a division of the observed super-Earth mass-radius diagram into five zones 
according to the inferred formation history.
\end{abstract}


\section{Introduction} \label{intro}

\begin{figure}
\begin{center}
\includegraphics[width=8cm]{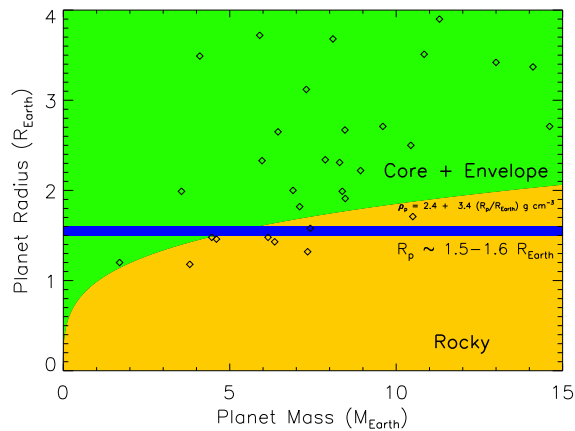}
\caption{The mass-radius diagram for observed exoplanets.
Among 65 exoplanets from the list of \citet{wm14}, 
33 exoplanets smaller than 4 $R_{\oplus}$ that have 2$-\sigma$ mass determinations are extracted,
following \citet{mwp14}.
In the data, planetary masses are obtained by radial velocity observations as well as transit timing variation (TTV)
while planet radii by transit observations such as the {\it Kepler} mission.
Coupling of these two observations implies that there may be a transition in planet radius 
($R_{tran} \simeq 1.5-1.6 R_{\oplus}$, also see the blue horizontal line)
above which most planets are very likely to be made of gaseous envelopes atop rocky cores,
below which most planets are very likely to be purely rocky \citep{r15}.
The dividing line between "Rocky" (the yellow regime) and "Core+Envelope" (the green regime) 
is given by an empirical linear density-radius relation \citep[][]{wm14,mwp14}.}
\label{fig1}
\end{center}
\end{figure}

The discovery of super-Earths has significantly enriched fundamental achievements of exoplanet observations. 
This was initially led by the radial velocity observations \citep[e.g.,][]{mq95,mb96,us07,mml11,fms14} 
and has subsequently been accelerated by the transit detections through {\it Kepler} mission \citep[e.g.,][]{bkb11,brb12,bbm14}.
Super-Earths that have $1-20 M_{\oplus}$\footnote{Exoplanets that have $10-20 M_{\oplus}$ are sometimes called as (hot) Neptunes, 
rather than super-Earths.} by a working definition are very peculiar 
in the sense that there is no such analogue in the solar system.
In fact, the mass range lies down between rocky planets like the Earth and the lower end of icy planets like the Neptune. 
The coupling of these two observations (the radial velocity and the transit)
reveals that the mean density of super-Earths is quite diverse \citep[e.g.,][, see Figure \ref{fig1}]{wm14,mwp14,jfr16,gcd16}.
These discoveries in turn stimulate active investigations of super-Earths, 
since their formation mechanisms are uncertain. 

Currently, three kinds of formation scenarios are predominantly examined (see Table \ref{table1}). 
The first one is the so-called "in situ" formation scenario \citep[e.g.,][]{ml09,cl13}.
In this picture, it is proposed that the observed super-Earths should have formed there 
through collisional growth of planetesimals and solids.
While the orbital distribution of the observed super-Earths allows construction of 
a minimum-mass extrasolar nebula (MMEN) model \citep{cl13},
which can be compatible with the famous minimum-mass solar nebula (MMSN) model \citep[e.g.,][]{h81},
the origin of high solid densities in the inner part of disks remains to be explored \citep[e.g.,][]{s14}.

\begin{table*}
\begin{minipage}{17cm}
\begin{center}
\caption{Currently proposed scenarios to form Super-Earths}
\label{table1}
\begin{tabular}{lccc} 
\hline 
                          & in-situ                                        &  embryo assembly                                       &  failed core (this study)                              \\ \hline 
Main mode of solid transport   &  radial drift of dust particles       &  type I migration ($M_p \la 1 M_{\oplus}$)  &  type II migration ($M_p \ga 2-3 M_{\oplus}$)  \\
Core formation  &  at $r < 1$ au in gas-rich disks  &  at $r < 1$ au in gas-poor disks                   &  at $r > 1$ au in gas-rich disks      \\ 
\hline 
\end{tabular}
\end{center}
\end{minipage}
\end{table*}

The other two kinds of scenarios require inward planetary migration that 
arises from tidal, resonant interactions between protoplanets and their gas disks \citep[e.g.,][]{kn12}.
Planetary migration can be classified into two modes, depending on the mass of migrators.
In general, type I migration is effective for low-mass planets such as terrestrial planets and cores of gas giants,
while type II migration becomes in action for massive planets such as Jovian planets 
that can open up a gap in their gas disks.

One of the remaining two scenarios is a scaled down version of gas giant formation \citep[e.g.,][]{abb06,mab09,rbl11,hp12,hp14}.
In this picture, planets form via the core accretion scenario, 
wherein cores of planets form initially, followed by the subsequent gas accretion onto the cores \citep[e.g.,][]{p96,il04i}.
The main difference with gas giant formation is that 
planetary cores that eventually become super-Earths cannot accrete the disk gas efficiently.
This inefficient gas accretion can be explained by the slow growth rate of planetary cores which should occur at the later stage of disk evolution. 
At that time, the formation of planetary cores will take a longer time and the core mass itself would become less massive, 
both of which lengthen the timescale of gas accretion onto the core and hence prevent the core from growing up to gas giants
even in gas disks.\footnote{Most of the currently existing studies implicitly assume that failed cores are dynamically isolated from others.
This may be possible due to their masses that are $\ga 5 M_{\oplus}$ as shown in this study.
Such planets may lead to dynamical clearing of the surrounding regions and hence suppress the subsequent giant impact stage (also see Section \ref{disc_2}).}
This formation mechanism thus can be referred to as a "failed core" scenario.

The second kind of scenarios with migration 
is an extension of rocky planet formation \citep[e.g.,][]{tp07,mn10,il10,hm12}.
In this scenario, 
planetary embryos that are massive enough to undergo rapid, inward type I migration but not to initiate efficient gas accretion, 
migrate toward the inner edge of disks and accumulate there.
Since gas disks are still present at that time, these embryos can establish a dynamically stable configuration.
This arises from efficient damping of the embryos' eccentricities by the disk gas, 
which eventually suppresses orbital crossing with each other.
Once the gas disks are gone, then these embryos collide together to form super-Earths.
An advantage of this scenario is that massive protoplanets that potentially accrete gas in the disks, 
can form only after gas disks disperse  significantly.
Therefore, there is no need of considering possibilities that 
protoplanets will grow rapidly via runaway gas accretion in gas disks.
In this paper, this scenario is called as "embryo assembly", 
since planets obtain most of their masses through collisions with other massive embryos and/or protoplanets.
 
As discussed above, one may classify the currently proposed scenarios for super-Earths,
depending on how solid materials are transported from the outer part to the inner part of disks,
and on when and where planetary cores are formed (see Table \ref{table1}).
It, however, should be noted that  
such a classification and each scenario are still under development.
In this paper, we will explore the the failed core scenario.
As already mentioned above,
the scenario essentially allows the formation of massive cores in gas disks, 
which can potentially trigger runaway gas accretion there and hence may end up with gas giants, rather super-Earths.
In addition, it is still a matter of debate how such massive cores are saved from rapid inward migration 
that leads to loss of the cores into the central star within the disk lifetime \citep[e.g.,][]{ward97,ttw02}.
Currently, it is well known that 
the direction of type I migration is quite sensitive to the disk structure \citep[e.g.,][]{ward97,ttw02,pm06,bm08,pbck09,hp10c}.
Nonetheless, the sensitivity causes further complexity of the migration,
because its systematic effect both on planetary formation and on the resultant populations of planets becomes unclear.
These considerations pose natural questions: 
what kind(s) of mechanisms would be needed to systematically understand the effect of planetary migration on the failed core scenario?
In the end, is the failed core scenario plausible to explain (at least in part) the population of observed super-Earths?

In this paper, we investigate the failed core scenario by taking account of planetary migration in detail 
and examine how important the scenario is to understand the observed properties of super-Earths.
We will explore the effect of gas accretion in a subsequent paper.
We, in particular, focus on planet traps - the stopping sites for rapid type I migration \citep{mmcf06,hp11}.
Combing the traps with the core accretion growth of planets, 
we have so far examined a number of characteristic features of the exoplanet observations 
such as the mass-period relation \citep{hp12}, the orbital distribution of exoplanets \citep{hp13a}, 
and the planet-metallicity relation \citep{hp14}.
Our previous studies suggest that 
planetary populations synthesized by the combination of the core accretion process with planet traps
are reasonably consistent with the trends of the exoplanet observations.
It is important that these results can be viewed as (partial) supportive evidence for the importance of planet traps.

We here apply our model to the population of super-Earths and 
discuss how significant planet traps are to form them.
We will demonstrate below that 
planet traps that can halt type I migration will play a crucial role in forming super-Earths,
as pointed out by our earlier work.
The movement of planet traps will terminate due to the dispersal of gas disks
as the traps arrive roughly at the distance of 1 au from the central star. 
This suggests that when planet traps act as the birth sites of super-Earths, 
the cores must have dropped out of their host traps and migrated to their current location possibly by type II migration.
The estimate of the gap-opening mass around $r=1$ au, therefore, provides a threshold value 
for trapped planets to be transported to the vicinity of the central star.
Under the presence of dead zones in our model that have $\alpha \simeq 10^{-4}$ using the $\alpha$-prescription \citep{ss73}, 
the gap-opening mass becomes about $3M_{\oplus}$ at the disk temperature of about 200 K at 1 au.
Furthermore, we perform a population synthesis analysis
and show that 
most of super-Earths generated by our model obtain the mass of about $4-5 M_{\oplus}$ or higher,
which is larger than the gap-opening mass at $r \simeq 1$ au.
We thus demonstrate that switching of migration modes can provide profound insights as to origins of close-in super-Earths.

The plan of this paper is as follows. 
In Section \ref{model}, a model used in this paper is described, 
wherein a recipe of planet traps and our population synthesis analysis are summarized.
In Section \ref{resu}, we present our results and 
examine the role of planet traps on the formation of super-Earths.
The results are derived from two distinct approaches: 
the one is to simply compute the closest orbital distance of planet traps from the central star, 
which is determined by disk evolution;
the other is to perform population synthesis calculations.
In Section \ref{disc}, we discuss other physical processes that may affect our findings, 
and provide some implications of our results for the mass-radius diagram of observed super-Earths.
Section \ref{conc} is devoted to the conclusion of this work.

\section{Models} \label{model}

We introduce a model used in this paper. 
Since a complete description of our model is given elsewhere \citep[e.g.,][]{hp12,hp14},
a summary of the model is provided here.
Parameters involved with our calculations are tabulated in Table \ref{table2}.
Note that all the parameters are identical to those used in \citet{hp14} (see their fiducial cases).

\begin{table*}
\begin{minipage}{17cm}
\begin{center}
\caption{List of parameters}
\label{table2}
\begin{tabular}{lll}
\hline
Symbol             &  Meaning                                                                                                                                     & Values           \\ \hline
Stellar parameters   &                                                                                                                   &                                       \\ \hline
$M_*$               &  Stellar mass                                                                                                                                & 1 $M_{\odot}$               \\
$R_*$               &  Stellar radius                                                                                                                                &  1 $R_{\odot}$              \\
$T_*$               &  Stellar effective temperature                                                                                                        & 5780 K                           \\ 
\hline
Disk model parameter       &                                                                                                        &                                       \\ \hline
$\tau_{vis}$      &   the viscous timescale (see equation (\ref{eq:mdot_exp}))                                                                 &   $10^6$ yr                  \\               
$\tau_{int}$      &   the initial time for starting computations (see equation (\ref{eq:mdot_exp}))                                    &   $10^5$ yr                  \\                       
$\Sigma_{A0}$  &  Surface density of active regions at $r=r_0$                                                                             & 20  g cm$^{-2}$ ($5 \leq \Sigma_{A0} \leq 50$) \\           
$s_A$                &  Power-law index of $\Sigma_A (\propto r^{s_A})$                                                                    & 3  ($1.5 \leq s_A \leq 6$)           \\
$\alpha_{A}$      &  Strength of turbulence in the active zone                                                                                 & $10^{-3}$  ($\alpha_A \leq 10^{-3}$)         \\        
$\alpha_{D}$     &  Strength of turbulence in the dead zone                                                                                   & $10^{-4}$ ($\alpha_D \leq 10^{-4}$)          \\  
$t$                     &  Power-law index of the disk temperature ($T \propto r^{t}$)                                                     & -1/2                               \\  
$f_{dtg}$           &   The dust-to-gas ratio at the solar metallicity  ($=\eta_{dtg} \eta_{ice}$)                             & $\simeq 1.8\times10^{-2}$      \\
$\eta_{dtg}$       &  The dust-to-gas ratio  within the ice line                                             & $\simeq 6\times10^{-3}$   \\
$\eta_{ice}$       &   A factor for increasing $f_{dtg}$ due to the presence of ice lines                            &  3                                    \\

\hline
Parameters for planetary growth      &                                                                                             &                                       \\ \hline
$M_{c,crit0}$    &   A free parameter regulating $M_{c,crit}$ (see equations(\ref{eq:m_ccrit}))                   &  $5M_{\oplus}$, $10M_{\oplus}$                \\
$(c,d)$               & A set of parameters for $\tau_{KH}$ (see Equation (\ref{eq:tau_KH}))               & (9,3)                                        \\
$f_{fin}$            &  Final mass of planets ($f_{fin} M_{gap}$)                                                                  & 10 ($>5$)                                  \\
\hline
Parameters for photoevaporative mass loss    &                                                                             &                                       \\ \hline
$p_1$               &   Parameter for $f_{lost} $ (see equation (\ref{eq:f_lost}))                                                           & 1.1 \\
$(p_2,p_3)$     &   A set of paramaters for $F_{th}$ (see equation (\ref{eq:f_th}))                                      &  (2.4,-0.7)               \\
$\epsilon$       &  Photoevaporative efficiency (see equation (\ref{eq:f_th}))                                                     &     0.1                                           \\   
\hline
Input parameters for the statistical analysis     &                                                                              &        \\   \hline
$\eta_{acc}$     &  A dimensionless factor for $\dot{M}$ (see equation (\ref{eq:mdot_exp}))                                         & $0.1 \leq \eta_{acc} \leq 10$  \\
$w_{mass}(\eta_{acc})$ & Weight function for $\eta_{acc}$ modeled by the Gaussian function                                &                                    \\
$\eta_{dep}$       &   A dimensionless factor for  $\tau_{dep}$ (see equation (\ref{eq:eta_dep}))                                    &  $0.1 \leq \eta_{dep} \leq 10$      \\                                                                                                                                              
$w_{lifetime}(\eta_{dep})$ & Weight function for $\eta_{dep}$ modeled by the Gaussian function                           &                                    \\
\hline
\end{tabular}

Note that the only free parameter in this work is $M_{c,crit0}$. 
For the other parameters, disk observations and theoretical modeling calculations provide certain values.
In addition, our previous study confirmed that the results are insensitive to a specific choice of the values for the parameters \citep{hp13a}.
\end{center}
\end{minipage}
\end{table*}

\subsection{Disk models} \label{disk_model}

We adopt a steady accretion disk model 
that still serves as an invaluable tool to characterize protoplanetary disks.
In the model, the disk accretion rate onto the central star is written as
\begin{equation}
 \dot{M}=3 \pi \nu \Sigma_g= 3 \pi \alpha c_s H \Sigma_g,
 \label{eq:mdot}
\end{equation}
where $\Sigma_g$, $\nu=\alpha c_s H$, $c_s$, and $H$ are the surface density, the viscosity, the sound speed, 
and the pressure scale height of gas disks, respectively. 
In order to parameterize the strength of disk turbulence, 
the famous $\alpha-$prescription is adopted \citep{ss73}.
This equation would be valid for the most region of protoplanetary disks. 

For the time evolution of disks, 
the following equation is used to change $\dot{M}$ with time ($\tau$) \citep{hp13a};
\begin{eqnarray}
 \label{eq:mdot_exp}
 \dot{M}(\tau) & \simeq & 3 \times 10^{-8} M_{\odot} \mbox{ yr}^{-1} \eta_{acc} 
                                                   \left( \frac{M_*}{0.5M_{\odot}}\right)^2 \\
              &    &  \times \left(1+ \frac{\tau}{ \tau_{vis}} \right)^{(-t+1)/(t-1/2)}  \exp \left( - \frac{\tau-\tau_{int}}{\tau_{dep}}  \right),  \nonumber                                   
\end{eqnarray}
where $\eta_{acc}$ is a parameter to scale $\dot{M}$ that will be discussed more in Section \ref{model_para} (also see Table \ref{table2}), 
$M_*$ is the mass of the central star,
$\tau_{vis}$ is the viscous diffusion timescale, $t$ is the exponent of the disk temperature ($T \propto r^t$), 
$\tau_{int}=10^5$ yr is the initial time of our computations,
and $\tau_{dep}$ is the disk depletion timescale,
which can be given as
\begin{equation}
\label{eq:eta_dep}
\tau_{dep} =  10^ 6 \eta _{dep}  \mbox{ yr}.
\end{equation}
This formulation is motivated by the current understanding of protoplanetary disks \citep[e.g.,][]{a11,wc11,hp13a,apa14,emn14};
in order to account for disk observations, 
disk evolution should be regulated by (at least) two physical processes, 
one of which is a slow ($\sim $ Myrs) diffusive process,
the other of which is a relatively rapid ($\sim 10^{4-5}$ yrs) dispersal process.
In the equation, the former one that is most likely governed by disk turbulence is represented by $\tau_{vis}$,
while the latter one that may be triggered by photoevaporation of disk gas is treated by $\tau_{dep}$.
This two-timescale behavior of disk dispersal has initially been proposed 
in the framework of photoevaporation-starved disk model \citep[e.g.,][]{cgs01,def09,oec11,oce12},
wherein $\tau_{dep}$ is involved with the photoevaporative outflow at $r =1-10$ au.
More recently, it has been suggested that magnetically driven winds from disk surfaces may play a similar role \citep[e.g.,][]{si09,smi10,bs13}.
Since the fundamental process of disk dispersal is still a matter of debate,
we adopt the above timescale approach.

As pointed out by \citet{hp13a}, 
equation (\ref{eq:mdot_exp}) also provides a behavior of $\dot{M}$ that is intermediate between the purely viscous evolution and 
the sharply truncated disk evolution that may occur due to a very efficient photoevaporation of gas disks (see their fig. 2).
Thus, our approach may be suitable for a statistical treatment of disk evolution 
in the sense that equation (\ref{eq:mdot_exp}) gives a median value of $\dot{M}$.

In summary, the disk accretion rate that is constant over the entire disk is given by equation (\ref{eq:mdot}), 
and its time evolution is characterized by two different timescales, $\tau_{vis}$ and $\tau_{dep}$ (see equation (\ref{eq:mdot_exp})).

\subsection{Planet traps} \label{pl_trap}

Planet traps that can halt rapid type I migration are one of the key ingredients in our model.
We briefly summarize their background and describe the recipe adopted in this paper.

As already mentioned in Section \ref{intro},
rapid type I migration arises from efficient transfer of the angular momentum between protoplanets and their natal gas disks \citep[e.g.,][]{gt80,ward97,ttw02}.
The transfer occurs only at the so-called Lindblad and corotation resonances, 
and the flow of transferred angular momentum depends deeply on types of resonances and the disk structure.
Since these resonant interactions inevitably lead to both gain and loss of the angular momentum from planets,
the net of transferred angular momentum determines the direction of planetary migration.
This in turn suggests that there is a possibility that the net torque can become zero at certain regions of disks,
which is the origin of planet traps.

The first recognition of planet traps was made by \citet{mmcf06}
in which 2D hydrodynamical simulations are performed.
Their simulations show that 
a cavity of gas disks can act as a barrier for rapid type I migration.
The barrier arises due to reversal of the torque balance near the cavity, which ends up with outward migration.
Since then, it has been proposed that a number of disk structures can become planet traps.
These include dead zones \citep[e.g.,][]{mpt07,hp10,rsc13} 
where magnetically induced turbulence is suppressed significantly due to the high column density of disks \citep[e.g.,][]{g96},
ice lines \citep[e.g.,][]{il08v,lpm10,hp11} 
at which the disk temperature becomes lower than the condensation temperature of certain molecules such as water, 
so that the molecules froze onto dust grains \citep[e.g.,][]{mdk11},
heat transitions \citep[e.g.,][]{hp11,kl12,yi12} 
where the main heat source of protoplanetary disks switches from viscous heating to stellar irradiation \citep[e.g.,][]{dccl98},
and opacity transitions \citep[e.g.,][]{lpm10,m11,hn12} at which the value of opacities changes significantly \citep[e.g.,][]{bl94}.
Thus, it is very likely that a number of planet traps can co-exist in single protoplanetary disks, 
and that these traps have characteristic orbital radii from the central star,
which are determined by their parental structures of disks 

In this paper, we follow a model of planet traps that was developed by \citet{hp11},
This model contains three kinds of traps: dead zone, ice line, and heat transition traps,
and the position of these traps is determined mainly by the disk accretion rate ($\dot{M}$).
We consider these traps because their parental disk structures can naturally be present as essential ingredients of protoplanetary disks.
Given that the time evolution of protoplanetary disks would be reasonably slow ($\sim$ Myrs),
these prominent disk structures may be able to exist over most of the disk lifetime, 
which can generate long-lived planet traps.
Also, it is assumed in this model that 
the effect of saturation is weak enough for planet traps to be effective when planets are less massive than the gap-opening mass.
This assumption is valid for most of the parameter space we consider, as shown below in Sections \ref{ch_mass} and \ref{resu_2}.

Here, we briefly summarize the orbital position of three traps \citep[see][for a complete derivation]{hp11}.
Assuming that dead zones are present in the inner part of disks, 
the effective value of $\alpha$ can be written as \citep[e.g.,][]{kl07,mpt09}
\begin{equation}
 \alpha = \frac{\Sigma_A \alpha_A+(\Sigma-\Sigma_A) \alpha_D}{\Sigma},
 \label{mean_alpha}
\end{equation}
where $\Sigma_A = \Sigma_{A0} (r/r_0)^{s_A}$ is the surface density of the active layer, 
and $\alpha_A$ and $\alpha_D$ are the strength of turbulence in active and dead layers, respectively. 
Then, the outer edge of dead zones that can serve as a planet trap can be given as \citep{hp11}
\begin{eqnarray}
 \label{eq:r_dz}
 \frac{r_{dz}}{r_0} & \simeq &  \left[ 71 \times \left( \frac{ \alpha_A+\alpha_D }{ 10^{-3}} \right)^{-1}
                               \left( \frac{ \Sigma_{A0} }{ 20 \mbox{ g cm}^{-2} } \right)^{-1} \right. \\ \nonumber
                      & \times &     \left.     \left( \frac{ H_0 }{0.1 \mbox{ au} } \right)^{-2}  
                               \left( \frac{ 2\pi / \Omega_0 }{1 \mbox{ yr}} \right)
                               \left( \frac{ \dot{M}/ 3\pi }{ 10^{-8} M_{\odot} \mbox{ yr}^{-1} } \right) \right]^{1/(s_A+t+3/2)},
\end{eqnarray}
where the jump in the surface density is established.
For an ice line trap, its position can be written as  
\begin{eqnarray}
 \label{eq:r_il}
 \frac{r_{il}}{r_0}  & \simeq & 14 \times \left( \frac{ \alpha_D }{ 10^{-4} } \right)^{-2/9}
                                              \left( \frac{T_{m,\mbox{H}_2\mbox{O}} (r_{il}) }{ 170 \mbox{ K} } \right)^{-8/3}   \\ \nonumber
                          &   \times       &            \left( \frac{ 2 \pi / \Omega_0}{ 1 \mbox{ yr}} \right)^{-2/3}
                                              \left( \frac{ \dot{M}/ 3\pi }{ 10^{-8} M_{\odot} \mbox{ yr}^{-1} } \right)^{4/9},
\end{eqnarray}
where viscous heating acts as the main heat source and achieves the sublimation temperature of water,
which is about 170K.
Finally, the position of heat transition traps is
\begin{eqnarray}
 \label{eq:r_ht}
 \frac{r_{ht}}{r_0}  & \simeq & 33 \times \left( \frac{ \alpha_A }{ 10^{-3} } \right)^{-14/45}
                                                \left( \frac{r_0}{ 1 \mbox{ au}} \right)^{4/5} 
                                             \left( \frac{2\pi / \Omega_0}{ 1 \mbox{ yr}} \right)^{- 6/5}  \\ \nonumber
                          &   \times       &  \left( \frac{R_*}{ 1 R_{\odot}} \right)^{-8/15}
                                             \left( \frac{T_*}{ 5800 \mbox{ K}} \right)^{-16/15} 
                                               \left( \frac{ \dot{M}/ 3\pi }{ 10^{-8} M_{\odot} \mbox{ yr}^{-1} } \right)^{28/45},
\end{eqnarray}
where the temperature set by viscous heating becomes equal to the temperature determined by stellar irradiation.
Note that the ice line trap will become active when 
\begin{equation}
 \frac{r_{il}}{r_{dz}} >  \left( \frac{H}{r}(r_{dz}) \frac{\alpha_A + \alpha_D}{\alpha_A - \alpha_D} \right)^{1/(s_A+t/2+1)}
 \label{eq:cond_il}
\end{equation}
while the heat transition trap will become effective when $r_{ht}>r_{dz}$.

\subsection{Characteristic masses at planet traps} \label{ch_mass}

Planet traps are one of the promising ideas to resolve the complexity of planetary migration.
It is nonetheless important to emphasize that the traps become effective only for planets with certain masses.
We identify what mass of planets can be captured at planet traps \citep[also see section 5 of][]{hp12}.

The lower limit of planet mass for planet traps to be active is set by the condition that 
the speed of type I migration should be faster than the moving rate of planet traps.
Then planets undergoing type I migration can catch up with the movement of planet traps.
By equating the type I migration timescale with the timescale regulating the movement of planet traps,
the following equation is obtained \citep[see][for a complete derivation]{hp12}:
\begin{equation}
 M_{mig,I} = \frac{H_p^2 M_*^2}{2 K_{mig} \Sigma_{g,p} r_p^4 \Omega_p \tau_{d}},
 \label{eq:M_typeI}
\end{equation}
where the suffix, $p$, denotes that physical quantities are evaluated at $r=r_p$, 
and $K_{mig}=1-10$, depending on the optical thickness of disks \citep{pbck09}.
Note that the moving rate of planet traps is determined by disk lifetimes 
($\tau_d$, see equations (\ref{eq:r_dz}), (\ref{eq:r_il}), and (\ref{eq:r_ht})).
For simplicity, we set that $\tau_{d} = \tau_{vis}$.
Thus, planets will be captured at planet traps when $M_p > M_{mig,I}$.
The upper limit is determined by the gap-opening mass ($M_{gap}$),
which can be written as \citep[e.g.,][]{lp93,mp06}
\begin{equation} 
 \label{eq:M_gap}
          M_{gap} =   \mbox{min}\left[ 3 \left( \frac{H}{r} \right)^3, \sqrt{40 \alpha  \left( \frac{H}{r} \right)^5} \right] M_*.                            
\end{equation}
This condition arises simply because planet traps are effective only for type I migrators.

It is also pointed out that 
the mass dependence of planet traps can originate from a unique feature of the corotation torque - saturation \citep{kl12,hn12,bcm13,crh14,dmk14}.
In principle, the corotation torque is characterized by the gas motion at the so-called horse-shoe orbit \citep[e.g.,][]{ward91,m02}.
In the full, non-linear picture, saturation occurs because gas parcels completing one or more libration cycles in the corotation region
exchange zero net angular momentum with planets \citep[e.g.,][]{pnk07}.
In protoplanetary disks, saturation can be avoided by viscous diffusion.
The diffusion restores the corotation torque by moving gas parcels off the librating orbits after a non-integer number of cycles.
When entropy-related corotation torque is considered, 
which was discovered numerically by \citet{pm06} 
and can play an important role in exciting planet traps \citep[e.g.,][]{lpm10,hp11,kl12},
another diffusion, also known as thermal diffusion, is required to inhibit saturation \citep[e.g.,][]{pbk10}.
The diffusion is involved with thermodynamics of disks,
which is poorly constrained in protoplanetary disks.

In this paper, we examine the effects of saturation by focusing on viscous diffusion,
since the diffusion is more fundamental for preventing saturation.
Based on the above argument, saturation can be avoided 
when the (local) viscous diffusion timescale ($\tau_{\nu}$) is shorter than the so-called libration timescale ($\tau_{lib}$)
which is involved with the gas motion across the horse-shoe orbit.
In the actual formula, the effect of saturation becomes weak enough for planet traps to be active when
\begin{equation}
 \label{eq:f_nu}
 \frac{f_{\nu} \tau_{\nu}}{\tau_{lib}} = \frac{f_{\nu} x_s^2/(3 \nu)}{8\pi r /(3 \Omega x_s)}<1,
\end{equation}
where $x_s$ is the width of the horse-shoe orbit, and a factor of $f_{\nu}(=0.55)$ is included following \citet{dmk14} \citep[also see][]{cpa16,apc16}.
Note that the value of $f_{\nu}$ is determined by fitting the results of analytical torque formula 
to those of numerical simulations done both by \citet{kbk09}  and by \citet{bk11}.
Since the best fit, which is $f_{\nu}=0.55$, is obtained where the net torque becomes zero (see figures 2 and 3 of \citet{dmk14}),
some degree of saturation is taken into account in equation (\ref{eq:f_nu}).
Adopting the results of numerical simulations \citep[e.g.,][]{mdk06,bm08,pbck09}, 
$x_s$ can be written as
\begin{equation}
x_s = 1.16 r \sqrt{ \frac{1}{\sqrt{\gamma}}\frac{M_p}{M_*} \frac{r}{H} },
\end{equation}
where $\gamma = 7/5$ is the adiabatic index.
As a result, planet traps become effective under the presence of (partial) saturation, when planets are less massive than
\begin{equation}
\label{eq:M_sat}
M_{sat}  = \left[  \frac{8 \pi \gamma^{3/4}}{(1.16)^3 f_{vis}} \alpha \left( \frac{H}{r} \right)^{7/2}  \right]^{2/3} M_*.
\end{equation}
It is important that $M_{sat} \approx M_{gap}$ as shown in Section \ref{resu_2}.
This indicates that in our framework, 
viscous diffusion is high enough to make saturation less effective for rapid type I migration.
Equivalently, planet traps become active when $M_p <  M_{gap} ( \approx M_{sat})$.
At the same time, the diffusion is low enough to allow planets to open up a gap in their gas disks
when their mass is larger than $M_{gap}$.
In other words, saturation becomes strong enough to null corotation torque in this mass range,
so that migration is regulated mainly by Lindblad torque.

In summary, planet traps are active when $M_{mig,I} < M_p <  M_{gap} ( \approx M_{sat})$.

\subsection{Planetary growth \& orbital evolution} \label{pl_growth}

We describe a prescription of both planetary growth and orbital evolution used in the model.

We adopt the core accretion scenario to form planets,
wherein planetary growth is characterized by two successive processes: 
the first step of forming planetary cores, and the second step of gas accretion on the cores 
\citep[e.g.,][]{p96,il04i,mab09,hp12,py14,lco14,pym15,lc15,lc16}.
For the core formation, the standard oligarchic growth is used 
to model how planetary cores will be built at planet traps through disk evolution.
More specifically, we start from embryos that have the mass of $\simeq 0.01 M_{\oplus}$,
and compute the mass evolution of such bodies 
based on the growth timescale ($\tau_{c,acc}$) given by \citet[see their equation (6)]{ki02}.
We have confirmed that a specific choice of the initial embryo mass does not affect our results very much.
The timescale is the outcome of more detailed N-body simulations 
which investigate how protoplanets will form through collisions among planetesimals.
This stage is referred to as oligarchic growth \citep[e.g.,][]{ki98,ki00,tdl03}.\footnote{
Note that oligarchic growth is a subsequent phase of runaway growth 
in which progenitors of oligarchs form quickly \citep[e.g.,][]{ws89,ki96}.} 
Since the timescale is a function of the solid density of disks ($\Sigma_d$),
we use the value of $\Sigma_d$ at planet traps that will change with time
as disks evolve.
More specifically,
$\Sigma_d$ at a dead zone, an ice line, and a heat transition trap can be given as below, 
respectively (also see equation (\ref{eq:mdot_exp}));
\begin{equation}
 \Sigma_{d,dz} \approx \frac{2 \dot{M} f_{dtg}}{3 \pi (\alpha_A+\alpha_D) H^2 \Omega}, 
\end{equation}
\begin{equation}
  \Sigma_{d,il} \approx \frac{\dot{M} f_{dtg}}{3 \pi  (\alpha_A+\alpha_D) H^2 \Omega},
\end{equation}
and
\begin{equation}
  \Sigma_{d,ht} \approx \frac{\dot{M} f_{dtg}}{3 \pi \alpha_A H^2 \Omega},
\end{equation}
where $f_{dtg}$ is the dust-to-gas ratio at the solar metallicity (see Table \ref{table2}).
Note that we have adopted that $\alpha \simeq \alpha_A+\alpha_D$ at ice line traps as dead zone traps.
This is because ice-coated dust grains that can be present at ice lines may absorb free electrons,
which can suppress MHD turbulence triggered by MRIs \citep[e.g.,][]{smun00,il08v}.

As time goes on, the core formation stage moves to the gas accretion one.
The classical, but milestone work by \citet{p96} proposes that 
gas accretion is composed of two distinct phases \citep[also see][]{hbl05,lhdb09,mbp10}: 
the one is a prolonged, slow gas accretion phase, so-called "phase 2";
the other is a rapid phase that starts when the envelop mass becomes comparable to the core mass,
which is referred to as the critical core mass \citep[$M_{c,crit}$, also see][]{m80,s82,bp86,r06}.
As already discussed in our earlier work \citep[see their section 2.1]{hp14},
the presence of the prolonged phase 2 is still a matter of debate \citep[also see][]{ine00,fbb07,si08}.
This is because the fundamental origin of the phase 2 is the continuous planetesimal accretion onto cores 
with a high accretion rate of $\dot{M}_c\sim 10^{-6} M_{\oplus}{\rm yr}^{-1}$ \citep{p96};
the infall of planetesimals releases their gravitational energy, 
which heats up gaseous envelopes surrounding cores and 
hence postpones the onset of rapid gas accretion onto the cores.
As pointed out by \citet{si08}, however, 
gravitational scattering caused by protoplanets pumps up the eccentricity of surrounding planetesimals.
Under the presence of the disk gas which can damp the planetesimals' eccentricities, 
the scattering leads to formation of a gap around the protoplanets in their planetesimal disk.
Consequently, the prolonged, efficient planetesimal accretion onto cores can be inhibited.
Taking into account the uncertainties in gas accretion processes, 
we adopt a conservative approach
wherein the critical core mass is used for the onset of gas accretion
and the Kelvin-Helmholtz timescale is utilized for regulating gas accretion.

In the actual formula, $M_{c,crit}$ is expressed as \citep[e.g.,][]{ine00}
\begin{equation}
 M_{c,crit} \simeq M_{c,crit0} \left( \frac{\dot{M}_c}{10^{-6}M_{\oplus} \mbox{ yr}^{-1}} \right)^{1/4}.
 \label{eq:m_ccrit}
\end{equation}
It is important to point out that the value of $M_{c,crit0}$ involves the grain opacity of accreting envelopes surrounding cores,\footnote{
Note that equation (\ref{eq:m_ccrit}) is derived 
under the assumption that gas supplies from the surrounding disk are constant.
As shown by \citet{lco14}, gas depletion that is significant in the later stage of disk evolution 
can become an important factor for regulating the value of $M_{c,crit}$.
This effect is not included in equation (\ref{eq:m_ccrit}).}
and is not a free parameter.
While the canonical value is $M_{c,crit0} \simeq 10 M_{\oplus}$ \citep[e.g.,][]{p96,ine00},
the recent studies suggest that $M_{c,crit0} < 10 M_{\oplus}$,
which indicates that grain growth can take place in planetary atmospheres \citep[e.g.,][]{mbp10,mka14,o14,hp14}.
Accordingly, we consider two values of $M_{c,crit0}$ in this paper: 
$M_{c,crit0} = 5 M_{\oplus}$, and $M_{c,crit0} = 10 M_{\oplus}$ (see Table \ref{table2}).
For the gas accretion, the Kelvin-Helmholtz timescale is used to model contraction of envelopes,
which is given as \citep[e.g.,][]{ine00}
\begin{equation}
 \tau_{KH} \simeq 10^{c} \mbox{ yr} \left( \frac{M_p}{M_{\oplus}} \right)^{-d}.
 \label{eq:tau_KH}
\end{equation}
While it is very likely that there are ranges in $c$ ($8 \la c \la 10$) and in $d$ ($2 \la d \la 4$), 
we use $c=9$ and $d=3$ in this paper \citep[e.g.,][]{il04i,hp14}.
Based on equation (\ref{eq:tau_KH}),
the gas accretion timescale varies with the computed mass ($M_{p}$) of planets.
For instance, $\tau_{KH} \simeq 8 \times 10^6 \rm yr$ when $M_{p}=5 M_{\oplus}$
while $\tau_{KH} \simeq 10^6 \rm yr$ when $M_{p}=10 M_{\oplus}$.

The orbital evolution of planets takes place simultaneously with planetary growth.
We consider two kinds of planetary migration in gas disks: trapped and type II migration \citep{hp12,hp13a}.
As discussed above, planet traps are effective for protoplanets that will undergo rapid type I migration.
Since the position of planet traps moves inward gradually through disk evolution,
protoplanets that are captured at planet traps will also spiral to the central star slowly 
according to the movement of their host traps (see equations (\ref{eq:r_dz}), (\ref{eq:r_il}), and (\ref{eq:r_ht})).
Note that the growth of all the protoplanets has been monitored,
and the protoplanets can follow the movement of their host trap only when $M_p > M_{mig,I}$.
The trapped migration would be valid until growing protoplanets reach the so-called gap-opening mass ($M_{gap}$)
and undergo type II migration \citep[e.g.,][]{lp86ii,npmk00,hi13}. 
When protoplanets obtain $M_{gap}$ (see equation (\ref{eq:M_gap})),
the tidal torque exerted by the protoplanets is large enough to open up a gap in their gas disks.
For this case, the migration rate can be determined by the local viscous diffusion rate ($\simeq f_{type II} \nu /r$),
where \citep[e.g.,][]{sc95,hi13}
\begin{eqnarray}
 \label{eq:M_typeII}
 f_{type II} & = & 1                          \mbox{ when }   M_p < 2 \pi \Sigma_g r^2(\equiv M_{mig,II}) \\ \nonumber
 f_{type II} & = & M_{mig,II}/M_p    \mbox{ when }   M_p > M_{mig,II}.
\end{eqnarray}

\subsection{Mass loss from planets} \label{mass_loss}

The main target of this paper is to focus on the failed core scenario and planet traps,
and examine the resultant population of super-Earths orbiting in the vicinity of the central star.
Nonetheless, it is interesting to consider the fate of fully formed planets.
This is because close-in planets are exposed to stellar illumination 
that is energetic enough to trigger mass loss from the planets \citep[e.g.,][]{kp83,lsr03,bsc04,vig10,lf13}.
To this end, we newly include the effect of photoevaporative mass loss from planets in our model.

We heavily rely on a model developed by \citet{lf13}. 
In the model, 
a fraction of the envelope mass ($f_{lost}$) that will be stripped away from planets via photoevaporation is 
described as a function of the core mass of planets ($M_c$),
which is given as 
\begin{equation}
\label{eq:f_lost}
f_{lost} = 0.5 \left( \frac{F_p}{F_{th}} \right)^{p1},
\end{equation}
where $F_p$ is the flux which planets receive from the central star, and 
\begin{equation}
\label{eq:f_th}
F_{th} = 0.5 F_{\oplus} \left( \frac{M_{c}}{M_{\oplus}} \right)^{p2} \left( \frac{\epsilon}{0.1} \right)^{p3}
\end{equation}
is the threshold flux to determine the photoevaporative mass loss, 
and $\epsilon$ is the photoevaporation efficiency. 
Following \citet{lf13}, we adopt that $p1 \simeq 1.1$, $p2 \simeq 2.4$, $p3 \simeq -0.7$, and $\epsilon = 0.1$.
The dependency on the core mass in equation (\ref{eq:f_th}) reflects the results of \citet{lf13} 
which show that
$M_c$ is one of the most fundamental parameters to scale $f_{lost}$ uniquely.

Based on the above two equations,
the value of $F_p$ and $M_c$ is needed to compute $f_{lost}$;
when $f_{lost}$ =1, the entire envelope is stripped away from planets.
Note that $F_{p}$ is a function of both the final position ($r_p$) of planets and the time when the planets arrive there.
We assume that stellar flux changes with time as $\propto \tau^{-1.23}$ \citep{rgg05,lfm12}.
All the input quantities ($M_c$, $M_p$, $r_p$ and $F_p$) are readily obtained in our model,
since the model can track down the full formation histories of planets growing at planet traps (see below).

It should be noted that the above analytical model is valid 
as long as the envelope mass is smaller than about 50 \% of the total planet mass (i.e., $M_c \ga 0.5 M_p$) \citep[][see their fig 7]{lf13}.
This is the outcome that, as envelopes grow in mass,
their self-gravity becomes more important such that photoevaporative mass loss is characterized by non-linear behaviors more.
The non-linearlity is not captured well by equations (\ref{eq:f_lost}) and (\ref{eq:f_th}).
In our population synthesis calculations, therefore, 
we take into account the photoevaporative mass loss only for planets that satisfy the condition that $M_c \ga 0.5 M_p$.

\subsection{Synthesizing planetary populations} \label{planet_pop}

Armed with the above processes and the related formulations,
we can compute evolutionary tracks of planets growing at planet traps in evolving protoplanetary disks \citep{hp12}.

Initially, embryos of planets are put at the position of planet traps in their disks.
As the disks evolve with time (see equation (\ref{eq:mdot_exp})),
planets grow in mass and move inwards, 
following the movement of planet traps when $M_p > M_{mig,I}$ (see equations (\ref{eq:r_dz}), (\ref{eq:r_il}), and (\ref{eq:r_ht})).
As found by \citet{hp12}, most of planets obtain the gap-opening mass (see equation (\ref{eq:M_gap}))
when planets already undergo gas accretion (see equation (\ref{eq:tau_KH})).
Planet traps are therefore effective both for the core formation stage and for the initial phase of gas accretion.
Type II migration takes place in the rest of the gas accretion stage.
The onset of gas accretion is determined by equation (\ref{eq:m_ccrit}).
Monitoring the core mass of planets that increases with time,
we find out when the computed core mass exceeds the critical value,
so that gas accretion begins.

Planetary growth via gas accretion will be terminated when planets achieve $f_{fin} M_{gap}$ with $f_{fin} =10$.
While gap formation via tidal torque and/or disk dispersal at the final evolution stage would determine the final mass of planets \citep[e.g.,][]{tt16},
both processes are still poorly constrained; 
considerable gas flow across a gap opened up by planets is observed in numerical simulations \citep[e.g.,][]{lsa99,ld06,tt16},
and it is far from a complete picture of disk dispersal (see Section \ref{disk_model}).
Under the circumstance, \citet{hp13a} perform a parameter study about $f_{fin}$, 
and show that the trend of the resultant planetary populations does not vary significantly when $f_{fin}> 5$ (see their section 7.3).
As a result, we adopt $f_{fin} =10$ in this paper.
We follow disk evolution until $\dot{M} = 10^{-14} M_{\odot} \mbox{ yr}^{-1}$ or $\tau=10^9$ yr.
Any kind of migration is stopped when migrators arrive at $r = 0.04$ au.
In other words, the final position of planets is either $r=0.04$ au 
or the location at which planets arrive when disk evolution is stopped.
Once planets arrive at their final position, 
photoevaporative mass loss is computed, following equations (\ref{eq:f_lost}) and (\ref{eq:f_th}). 
In our model, planet formation can take place simultaneously at three planet traps in single disks.
In this paper, we neglect gravitational interactions that can potentially occur among planets growing at three different traps,
and will include it in a subsequent paper (also see Section \ref{disc_2}).

\begin{table*}
\begin{minipage}{17cm}
\begin{center}
\caption{Preferred places of observed exoplanets}
\label{table3}
\begin{tabular}{lcc} 
\hline 
Definition$^1$                                   &  Mass range ($M_{\oplus}$)   &  Semimajor axis range (au)          \\ \hline 
Hot Jupiters  (Zones 1 and 2)$^2$   &  30 $< M_p<$ $10^4$            &  0.01 $< r_p <$ 0.5                             \\ 
Exo-Jupiters  (Zone 3)$^2$              &  30 $< M_p<$ $10^4$            &  0.5   $< r_p <$ 10                             \\ 
Low-mass planets (Zone 5)$^2$      &  1 $< M_p<$ 30                      &  0.01   $< r_p <$ 0.5                        \\
\hline 
\end{tabular}

$^1$ the same definitions as \citet{hp14}

$^2$ the definition made in \citet{hp13a}

\end{center}
\end{minipage}
\end{table*}

Utilizing the above technique to compute the tracks,
we have developed our version of population synthesis model \citep{hp13a}.
In the model, there are three key ingredients.
First, fully formed planets are categorized into three clasess, 
based on their masses and orbital distances (see table \ref{table3}).
The classification is originally motivated by the exoplanet observations; 
exoplanets may have preferable locations in the mass-semimajor axis diagram 
at which the observed exoplanets are densely populated \citep[e.g.,][]{cl13,hp13a}.
While we can add more classes into table \ref{table3},
we simply follow \citet{hp14}, 
since the classification does not affect our conclusions significantly.
Second, a large number of tracks ($N_{int}=300$ in total) are computed 
for a given set of the disk mass ($\eta_{acc}$) and lifetime ($\eta_{lifetime}$) parameters 
(see equation (\ref{eq:mdot_exp}), also see Table \ref{table3}).
We have confirmed that the total number of tracks is large enough for the results to converge \citep{hp13a}.
This can be viewed as one of the confirmations that within our framework,
the calculations cover all the possibilities of planet formation that can occur at every stage of disk evolution.
Finally, we compute planet formation frequencies (PFFs).
This can be done by two successive processes.
The initial process is to count all the end points of tracks that are eventually located at each zone ($N\mbox{(Zone i, } \eta_{acc}, \eta_{dep}$)).
Note that Zone $i$ is defined, following \citet{hp13a} (see Table \ref{table3}).
Since such counting is done for a given set of both the disk mass ($\eta_{acc}$) and lifetime ($\eta_{lifetime}$) parameters,
the second process is to change the value of both $\eta_{acc}$ and $\eta_{dep}$, 
and sum up the resultant $N\mbox{(Zone i, } \eta_{acc}, \eta_{dep}$)
with weight functions of these two parameters ($w_{mass}$ and $w_{lifetime}$ for $\eta_{acc}$ and $\eta_{dep}$, respectively).
Mathematically, the process can be written as
\begin{eqnarray}
 \label{pfr}
 \mbox{PFFs(Zone i)}   & \equiv  &  \\  \nonumber   
           \sum_{\eta_{acc}} \sum_{\eta_{dep}} & w_{mass}(\eta_{acc}) & w_{lifetime}(\eta_{dep}) \\  \nonumber        
          & \times &   \frac{N\mbox{(Zone i, } \eta_{acc}, \eta_{dep})}{N_{int}}.                           
\end{eqnarray}
We adopt $w_{mass}$ and $w_{lifetime}$ such that 
the observations of protoplanetary disks are reproduced reasonably well \citep{hp13a}.

In addition to the PFFs, we also compute the following two quantities;
the one is the statistically averaged value of the gap-opening mass ($M_{gap}$),
which is defined as
\begin{eqnarray}
 \label{M_gap_ave}
\braket{ M_{gap}\mbox{(Zone i)} }  & \equiv  &   \\  \nonumber   
           \sum_{\eta_{acc}} \sum_{\eta_{dep}} & w_{mass}(\eta_{acc}) & w_{lifetime}(\eta_{dep}) \\  \nonumber               
          & \times &   \braket{M_{gap}(\mbox{Zone i, } \eta_{acc}, \eta_{dep})},           
\end{eqnarray}
where $\braket{M_{gap}(\mbox{Zone i, } \eta_{acc}, \eta_{dep})}$ is the averaged gap-opening mass of planets 
that eventually fill out Zone i for a certain set of $\eta_{acc}$ and $\eta_{dep}$.
The other is the statistically averaged value of $M_{c,crit}$ (see equation (\ref{eq:m_ccrit})),
which can be written as
\begin{eqnarray}
 \label{M_ccrit_ave}
\braket{ M_{c,crit}\mbox{(Zone i)} }  & \equiv  &   \\  \nonumber   
           \sum_{\eta_{acc}} \sum_{\eta_{dep}} & w_{mass}(\eta_{acc}) & w_{lifetime}(\eta_{dep}) \\  \nonumber               
          & \times &   \braket{M_{c,crit}(\mbox{Zone i, } \eta_{acc}, \eta_{dep})},           
\end{eqnarray}
where $\braket{M_{c,crit}(\mbox{Zone i, } \eta_{acc}, \eta_{dep})}$ is the averaged $M_{cirit0}$ of planets
that eventually end up in Zone i for a given set of $\eta_{acc}$ and $\eta_{dep}$.
These two quantities are important to analyze the results of our population synthesis calculations (see below).

\subsection{Model parameters} \label{model_para}

Parameters involved in our population synthesis calculations can be classified into five categories (see Table \ref{table2}):
the stellar parameters, the disk model ones, the parameters for planetary growth, those for photoevaporative mass loss, 
and the input parameters for the statistical analysis.
Both the stellar parameters and the input parameters for the statistical analysis 
are central in synthesizing planetary populations as well as 
in qualitatively reproducing the exoplanet observations \citep[e.g.,][]{il04i,mab09,hp13a}.
It is important that they are essentially determined by the observations of stellar and disk populations.
Since the main scope of this paper is super-Earths orbiting around G stars,
we simply pick up stars of $M_* = 1 M_{\odot}$.

For the disk parameters, there are many parameters that are not constrained tightly.
This is particularly valid for the quantities related to dead zones.
\citet{hp13a} have therefore undertaken a parameter study about them,
and confirmed that the resultant planetary populations do not change very much 
when $5 \leq \Sigma_{A0} \leq 50$, $1.5 \leq s_A \leq 6$, $\alpha_A \leq 10^{-3}$, and $\alpha_D \leq 10^{-4}$ (also see Table \ref{table2}).
In addition, there are model parameters to regulate the efficiency of photoevaporative mass loss.
These parameters are determined by a fitting to the results of more detailed calculations \citep{lf13},
and can therefore be regarded that given values are constrained reasonably well.
Finally, there are parameters that belong to the category of planetary growth.
These parameters that involve the efficiency of gas accretion onto planetary cores,
can also be derived from comparison with more detailed calculations.
For $f_{fin}(=10)$, we rely on the results of \citet{hp13a} as discussed in Section \ref{planet_pop}.
For the value of $(c,d)$, we simply follow \citet{il04i} which suggest that the results are not affected substantially 
by changing them within $8 \la c \la 10$ and $2 \la d \la 4$.
In this paper, therefore, we treat $M_{c,crit0}$ as the only tunable parameter.
As discussed below, we examine how important the value of $M_{c,crit0}$ is to determine the resultant PFFs.

In summary, there are many parameters in population synthesis calculations.
Nonetheless, we can demonstrate how insensitive the results are to most of the parameters.
Eventually, $M_{c,crit0}$ is the only free parameter in this work.
 
\section{Results} \label{resu}

We present the results that are derived from two different approaches. 
In the first approach, we compute the position of planet traps that is determined by the disk accretion.
We also calculate the gap-opening mass ($M_{gap}$) that is estimated from the position of planet traps.
These results are obtained, in order to demonstrate how important type II migration is to form close-in super-Earths in our model.
In the second approach, we perform population synthesis calculations to develop a more quantitative analysis.
More specifically, we derive the minimum mass of planets that can eventually be categorized as close-in super-Earths in our model. 
In addition, we investigate how such a value of planet mass is determined by focusing on planetary migration as well as $M_{gap}$
in the framework of the failed core scenario.

\subsection{Positions of planet traps} \label{resu_1}

\begin{figure*}
\begin{minipage}{17cm}
\begin{center}
\includegraphics[width=8cm]{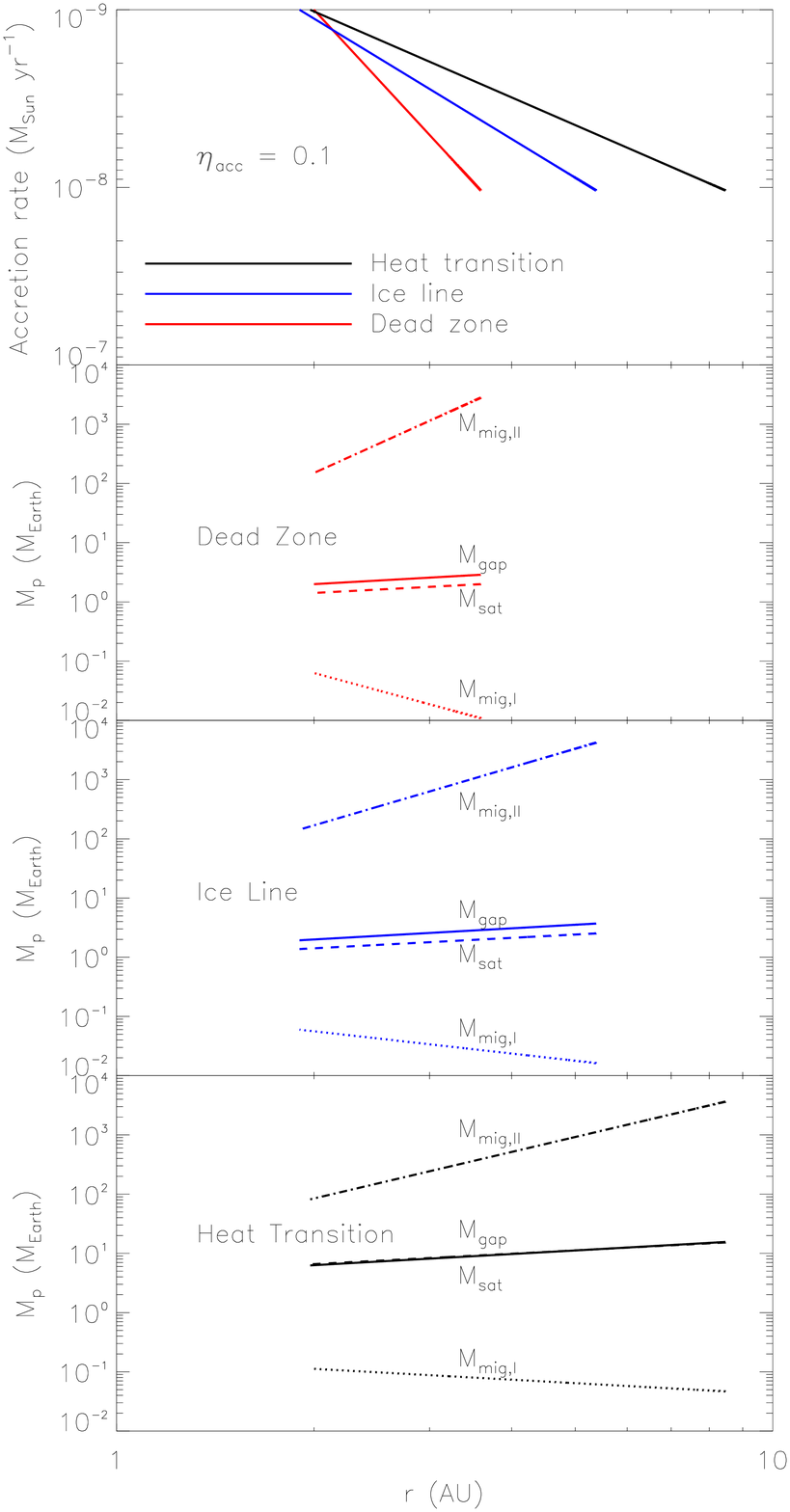}
\includegraphics[width=8cm]{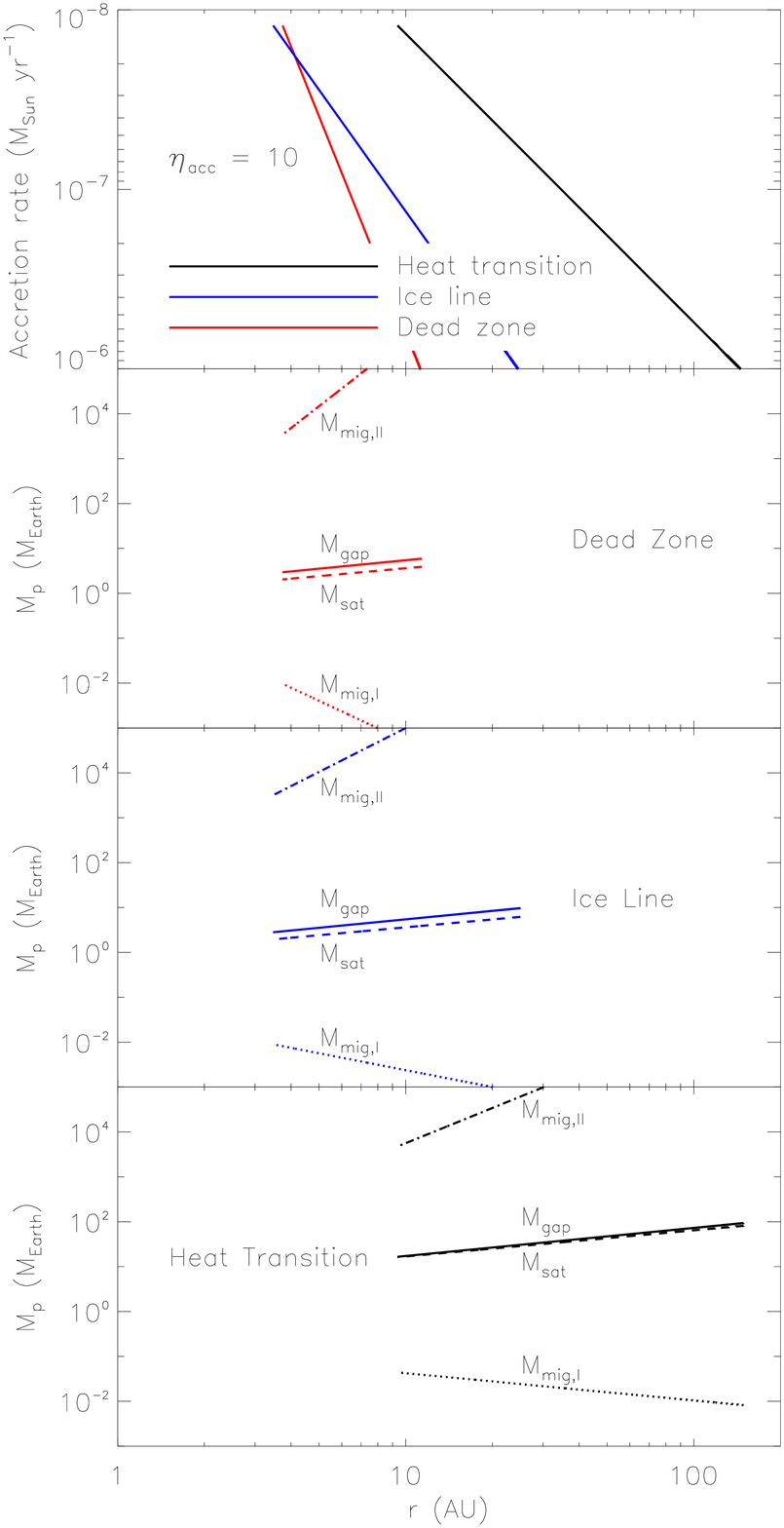}
\caption{The movement of planet traps and the corresponding values of $M_{mig,I}$, $M_{sat}$, $M_{gap}$, and $M_{mig,II}$.
The results for the case that $\eta_{acc}=0.1$ are presented on the left panel,
while those for the case that $\eta_{acc}=10$ are on the right one.
On top, the position of a dead zone trap is denoted by the red line, that of an ice line trap is by the blue line,
and that of a heat transition trap is by the black line.
From second to bottom, the resultant values of $M_{mig,I}$ (the dotted line), $M_{sat}$ (the dashed line), 
$M_{gap}$ (the solid line), and $M_{mig,II}$ (the dash-dotted line) are shown for a dead zone, an ice line, and a heat transition trap, resepctively.
At the early stage of disk evolution, three planet traps spread out widely.
As $\dot{M}$ decreases with time, the mutual distance between planet traps shrinks.
When $\tau=10^7$ yr or $\dot{M} = 10^{-9} M_{\odot} \mbox{ yr}^{-1}$ 
at which gas disks can severely be depleted due to photoevaporation,
the traps end up at $r \ga 1$ au.
Importantly, the results suggest that trapped planetary cores should drop out of their host trap, in order to be located within $r \simeq 1$ au.
Such dropping out can occur when protoplanets obtain $M_{gap}$, and hence can undergo type II migration.
We confirm that planet traps are effective for planets with the mass of $> M_{mig,I}$ and 
that the effect of (partial) saturation is weak enough in the trapped migration regime in our model since $M_{sat} \approx M_{gap}$.
Our results also show that type II migration can surely occur for planets with the mass of $> M_{gap}$,
because $M_{mig,II} > M_{gap}$.
We find that $M_{gap}$ is determined by the viscosity condition ($\sqrt{40 \alpha (H/r)^5}$) at all the traps
for these two cases of $\eta_{acc}$ (see equation (\ref{eq:M_gap})).
The results imply that the minimum mass of planets that can be transported from $r > 1$ au to $r < 1$ au via type II migration 
should be larger than $\sim 2-3 M_{\oplus}$.}
\label{fig2}
\end{center}
\end{minipage}
\end{figure*}

We first examine how the position of planet traps evolves with time.
Figure \ref{fig2} (top) shows their resultant behavior.
We compute the position of dead zone, ice line, and heat transition traps 
using equation (\ref{eq:r_dz}), (\ref{eq:r_il}), and (\ref{eq:r_ht}), respectively.
In the plot of the disk accretion rate ($\dot{M}$) as a function of the position of planet traps,
the position is determined uniquely by $\eta_{acc}$ (not by $\eta_{dep}$).
As a result, we here consider two extreme cases of $\eta_{acc}$ (i.e., $\eta_{acc}=0.1$ and $\eta_{acc}=10$) 
that  are presented on the left and the right panel, respectively.
Note that while population synthesis calculations will continue 
until $\dot{M} = 10^{-14} M_{\odot} \mbox{ yr}^{-1}$ or $\tau=10^9$ yr (see Section \ref{planet_pop}),
we here stop the calculations when $\dot{M} = 10^{-9} M_{\odot} \mbox{ yr}^{-1}$ or $\tau=10^7$ yr.
This is because disk observations suggest that these values are more relevant to the lifetime of gas disks \citep[e.g.,][]{wc11}.

As found by \citet{hp11}, 
Figure \ref{fig2} (top) shows that a dead zone trap is located in the innermost (see the red line), 
an ice line trap is in the middle (see the blue line), and a heat transition trap is in the outermost (see the black line).
When $\dot{M}$ is high, that is, disks are at the early evolution stage,
three traps distribute around $r \simeq 3-10$ au for the case that $\eta_{acc} =0.1$.
As time goes, the value of $\dot{M}$ decreases, and hence their position becomes much closer to the central star.
It is of fundamental importance that for the case that $\eta_{acc} =0.1$, 
all the traps arrive at $r \simeq 1$ au when the value of $\dot{M}$ is low enough to be subject to photoevaporation of gas disks.
For the case that $\eta_{acc} =10$, we obtain the qualitatively similar results (see the right panel of Figure \ref{fig2}).
The noticeable quantitative difference is that the position of planet traps spreads out over larger disk radii;
at the early stage of disk accretion,
a dead zone trap is located at $r \simeq 10$ au, an ice line trap is at $r \simeq 20$ au, and a heat transition trap is beyond $r \simeq 100$ au.
We emphasize that even for this case, the final position of planet traps is well beyond $r=1$ au.

Thus, our results suggest that, provided that planetary cores form predominantly at planet traps in gas disks,
there is the minimum orbital distance within which trapped cores cannot be transported by the movement of planet traps.
In other words, type II migration should come into play for the cores to be located within $r=1$ au (see below).

\subsection{Characteristic masses at planet traps} \label{resu_2}

We are now in a position to demonstrate how crucial type II migration is to form close-in (low-mass) planets in our model.
To proceed, we calculate the values of $M_{mig,I}$, $M_{sat}$, $M_{gap}$, and $M_{mig,II}$, following the movement of planet traps
(see equations (\ref{eq:M_typeI}), (\ref{eq:M_sat}), (\ref{eq:M_gap}), and (\ref{eq:M_typeII})).
As discussed below, the gap-opening mass ($M_{gap}$) is important 
to estimate the minimum mass of planets that would populate in the vicinity of the central star.

Figure \ref{fig2} (from second to bottom) shows 
the resultant values of $M_{mig,I}$, $M_{sat}$, $M_{gap}$, and $M_{mig,II}$ 
at the position of a dead zone trap, an ice line trap, and a heat transition trap, respectively.
The dotted line denotes the behavior of $M_{mig,I}$, the dashed line is for $M_{sat}$,
the solid line is for $M_{gap}$, and the dash-dotted line is for $M_{mig,II}$ at each trap.
As already demonstrated by \citet{hp12} (see their Appendix A),
the value of $M_{mig,I}$ increases as planet traps move inwards.
This is simply because $M_{mig,I}$ is a decreasing function of $r$ (see equation (\ref{eq:M_typeI})).
It is important that the value of $M_{mig,I}$ is very low,
so that once protoplanets obtain the mass of $\la 0.1 M_{\oplus}$,
the protoplanets can catch up with the inward movement of their host trap.
Thus, planet traps can become effective for planets with the mass of $> 0.1 M_{\oplus}$.
For the upper mass limit, the effectiveness of planet traps can be determined either by the gap-opening mass ($M_{gap}$, see equation (\ref{eq:M_gap}))
or by the effect of saturation when corotation torque plays an important role in generating the traps (see equation (\ref{eq:M_sat})).
We find that $M_{gap} \ga M_{sat}$ for dead zone and ice line traps and $M_{gap} \approx M_{sat}$ for a heat transition trap.
Quantitatively, the maximum differences between $M_{gap}$ and $M_{sat}$ are 
30 \% at a dead zone trap, 31 \% at a ice line trap, and 4 \% at a heat transition trap for the case of $\eta_{acc}=0.1$,
and 34 \% at a dead zone trap, 36 \% at a ice line trap, and 12 \% at a heat transition trap for the case of $\eta_{acc}=10$.
Such small differences can allow one to assume that 
the upper limit of planet mass for planet traps to be active can be given as $M_{gap}$ (rather than $M_{sat}$).
This assumption would be reasonable in our model because the effect of (partial) saturation would be very weak when $M_p < M_{sat}$.
As a result, protoplanets will be captured at their host trap in this mass range.
As planets grow in mass, they exceeds the value of $M_{sat}$.
Then, Lindblad torque is the main driver to regulate planetary migration.
At a heat transition trap, switching of migration modes from trapped I to type II migration occurs 
immediately after protoplanets achieve $M_{gap}$ (since $M_{gap} \approx M_{sat}$).
For dead zone and ice line traps, there are differences between $M_{gap}$ and $M_{sat}$,
but the differences are small.
Consequenly, protoplanets that have the mass of $> M_{sat}$ may undergo the standard inward type I migration due to Lindblad torque
after the effect of saturation becomes strong enough to make planet traps ineffective.
Nonetheless, they shift to the type II regime soon after dropping out of their host trap.
Thus, it would be reasonable to assume that protoplanets will follow the movement of planet traps 
when $M_{mig,I}< M_p < M_{gap}$.
In other words, the gap-opening mass and the subsequent type II migration are crucial 
to quantify the minimum mass of planets orbiting in the vicinity of the central star.

We now discuss the resultant behavior of $M_{gap}$ (see Figure \ref{fig2}). 
We find that the viscosity condition ($\sqrt{40 \alpha (H/r)^5}$) provides a lower value for all the three traps in both the cases of $\eta_{acc}$.
As a result, the trajectory of $M_{gap}$ in the mass-semimajor axis diagram has the same slope at all the traps.
Note that the value of $M_{gap}$ at a dead zone trap is exactly identical to that at an ice line trap.
This arises from the assumption that $\alpha \simeq \alpha_D$ at both dead zone and ice line traps at the disk midplane.
Our results also show that $M_{gap} \simeq 2-3 M_{\oplus}$ 
when both dead zone and ice line traps arrive at the smallest orbital distance.
We obtain similar results for the cases that $\eta_{acc}=0.1$ and that $\eta_{acc}=10$.
For the heat transition trap, the value of $M_{gap}$ at the end stage of disk evolution becomes higher 
than that at the dead zone and ice line traps for these two cases of $\eta_{acc}$.
It is important to confirm whether or not protoplanets dropping-out from their host trap can really undergo type II migration.
For this purpose, $M_{mig,II}$ is also plotted in Figure \ref{fig2} (see equation (\ref{eq:M_typeII})).
We find that the resultant value of $M_{mig,II}$ is much higher than $M_{gap}$ at all the traps, even in the later stage of disk evolution.

In summary, our results suggest that inward type II migration would be a vital mechanism 
to transport protoplanets formed at planet traps from $r >1$ au to $r < 1$ au,
and imply that the mass of planets should exceed $\sim 2-3 M_{\oplus}$ in order for them to be finally located in the vicinity of the central star.

\subsection{Planet formation frequencies} \label{resu_3}

As discussed above, our results suggest that 
the gap-opening mass and the resultant switching of migration modes are important to distribute planets in the vicinity of the central star,
under the action of planet traps.
Here, we present the results that are derived from the second approach
in which population synthesis calculations are carried out.
Since the calculations fully track down the evolution history of planets both in mass and in orbital radius,
we can now investigate how the above results are related to the formation of close-in super-Earths.
More specifically, we can quantitatively assess the importance of planet traps on such planets.
Before addressing these points,
we first discuss the resultant planet formation frequencies (PFFs) that are summarized in Table \ref{table4}.
Note that the total PFFs are not 100 \%.
This occurs because we count only planets that end up in the three regimes under consideration (see Table \ref{table3}).

\begin{table*}
\begin{minipage}{17cm}
\begin{center}
\caption{The resultant PFFs and other key quantities}
\label{table4}
\begin{tabular}{c|cccc|cc}
\hline
$M_{c,crit0}$          &   Hot Jupiters   &  Exo-Jupiters   & Low-mass planets  & Total        & $\braket{ M_{gap} }^a $ & $\braket{ M_{c,crit} }^a $ \\ \hline 
$5 M_{\oplus}$       &   7.9 \%            & 25.4 \%            & 10.0 \%                   & 43.3  \%  & 3.7 $M_{\odot}$        &  2.6 $M_{\odot}$ \\
$10 M_{\oplus}$     &   9.9 \%            & 30.1 \%            & 4.9 \%                     & 44.9 \%   & 3.4 $M_{\odot}$        &  2.8 $M_{\odot}$  \\
\hline
\end{tabular}

$^a$ the value computed for planets that eventually fill out the Low-mass planet regime
\end{center}
\end{minipage}
\end{table*}

Our results show that the PFF of exo-Jupiters is the highest
when planets form via the core accretion process coupled with planet traps.
As discussed intensively in \citet{hp13a}, 
this becomes possible due to the presence of planet traps \citep[also see][]{hp12};
cores of gas giants can form at $r \simeq 1-10$ au 
when the disk mass is about a few times higher than the MMSN \citep[e.g.,][]{il04i,hbl05,tmr08}.
In general, such cores will experience rapid (inward) type I migration and have difficulty in growing up to gas giant there.
As shown in Figure \ref{fig2}, however,
planet traps can sweep up such a region of protoplanetary disks over a long time ($\simeq$ Myr).
This leads to a situation that embryos of protoplanets that can otherwise plunge into the central star within disk lifetimes 
can be captured at planet traps and will distribute there.
Consequently, the presence of planet traps makes it possible to efficiently form exo-Jupiters in our model.
The results, contrastingly, show that the PFF of hot Jupiters is much lower than that of exo-Jupiters (see Table \ref{table4}).
This is the outcome that, in order to fill out the regime of hot Jupiters,
a long disk lifetime (i.e., a large value of $\eta_{dep}$) is needed.
Then, gas giants formed at $r \ga 1$ au can have enough time to move to the vicinity of the central star via type II migration.
Since the current disk observations suggest that long-lived gas disks are minor \citep[e.g.,][]{wc11},
the population of hot Jupiters becomes small.
These behaviors are valid for both the cases that $M_{c,crit0}=5 M_{\oplus}$ and $M_{c,crit0} =10 M_{\oplus}$.
Thus, we demonstrate that the resultant PFFs of jovian planets are consistent with the trend of exoplanet observations 
which indicate that gas giants are densely populated around $r \simeq 1$ au with fewer hot Jupiters.

Table \ref{table4} also shows that 
the population of Low-mass planets generated by our model is the second dominant for the case that $M_{c,crit0} = 5M_{\oplus}$.
These planets are central in this paper, because they are essentially comparable to observed super-Earths (see Table \ref{table3}).
We first emphasize that these planets are regarded as "failed cores" in our calculations.
This is because we model planet formation that can occur only in gas disks.
In other words, planets that fill out the regime of Low-mass planets have solid cores that can potentially grow up to gas giants.
Nonetheless, such cores cannot experience significant gas accretion.
As demonstrated clearly in \citet[see their fig 3]{hp13a}, this occurs due to the formation timing; 
most of the cores that eventually grow up to super-Earths start forming at the late stage of disk evolution. 
At that time, the solid density ($\Sigma_d$) is low,
and hence the core formation timescale becomes long.
This slow growth of cores delays the onset of gas accretion.
Furthermore, the cores themselves become less massive for this case \citep[see][]{hp14}.
As a result, the subsequent gas accretion becomes prolonged.
In fact, we find that $\braket{ M_{c,crit} } \la 3 M_{\oplus}$ for planets in the Low-mass regime (see Table \ref{table4}).
Based on such a value, the Kelvin-Helmholtz timescale becomes $\simeq 3.7 \times 10^7$ yr (see equation (\ref{eq:tau_KH})),
which obviously exceeds the disk lifetime. 
In addition, we have confirmed that both dead zone and heat transition traps contribute to the PFFs of the Low-mass planets equally
as shown in \citet[see their table 4]{hp13a}.
It is important that these arguments are maintained for both the cases that $M_{c,crit0}=5 M_{\oplus}$ and $M_{c,crit0} =10 M_{\oplus}$.
Thus, our results show that planets in the Low-mass planets region are made of solid cores surrounded by gaseous envelopes.

We then must admit that the resultant PFFs of Low-mass planets are not in a good agreement with the trend of the current exoplanet observations.
This is because exoplanet observations indicate so far that close-in super-Earths are the most abundant \citep[e.g.,][]{hmj10,mml11,ftc13}.
We consider that this difference simply suggests that there are a number of mechanisms to form super-Earths.
As described in Section \ref{intro} (see Table \ref{table1}), 
the failed core scenario may be one of them to fully reproduce the diversity of observed super-Earths.
In the following sections, we examine how useful the failed core scenario is to understand some features of close-in super-Earths.

\subsection{Planets in the Low-mass regime}

As pointed out in Sections \ref{resu_1} and \ref{resu_2},
the movement of planet traps shuts off at $r \ga 1$ au due to the dispersal of gas disks,
and hence type II migration is needed for forming planets to distribute within $r \la 1$ au.
Figure \ref{fig2} suggests that planets more massive than $\sim 2-3 M_{\oplus}$ 
can open up gaps in the disk and drift inward via type II migration.
On the other hand, our population synthesis calculations confirm above that the coupling of the core accretion scenario with planet traps 
can generate a considerable fraction of planets that can be viewed as close-in super-Earths (see Table \ref{table4}).
Here, we focus on the population of Low-mass planets that is synthesized by our calculations,
and investigate what mass of planets formed in our model can eventually fill out the super-Earth regime.
This enables us to explore what is the relationship between such planetary masses and type II migration.

\begin{figure*}
\begin{minipage}{17cm}
\begin{center}
\includegraphics[width=8cm]{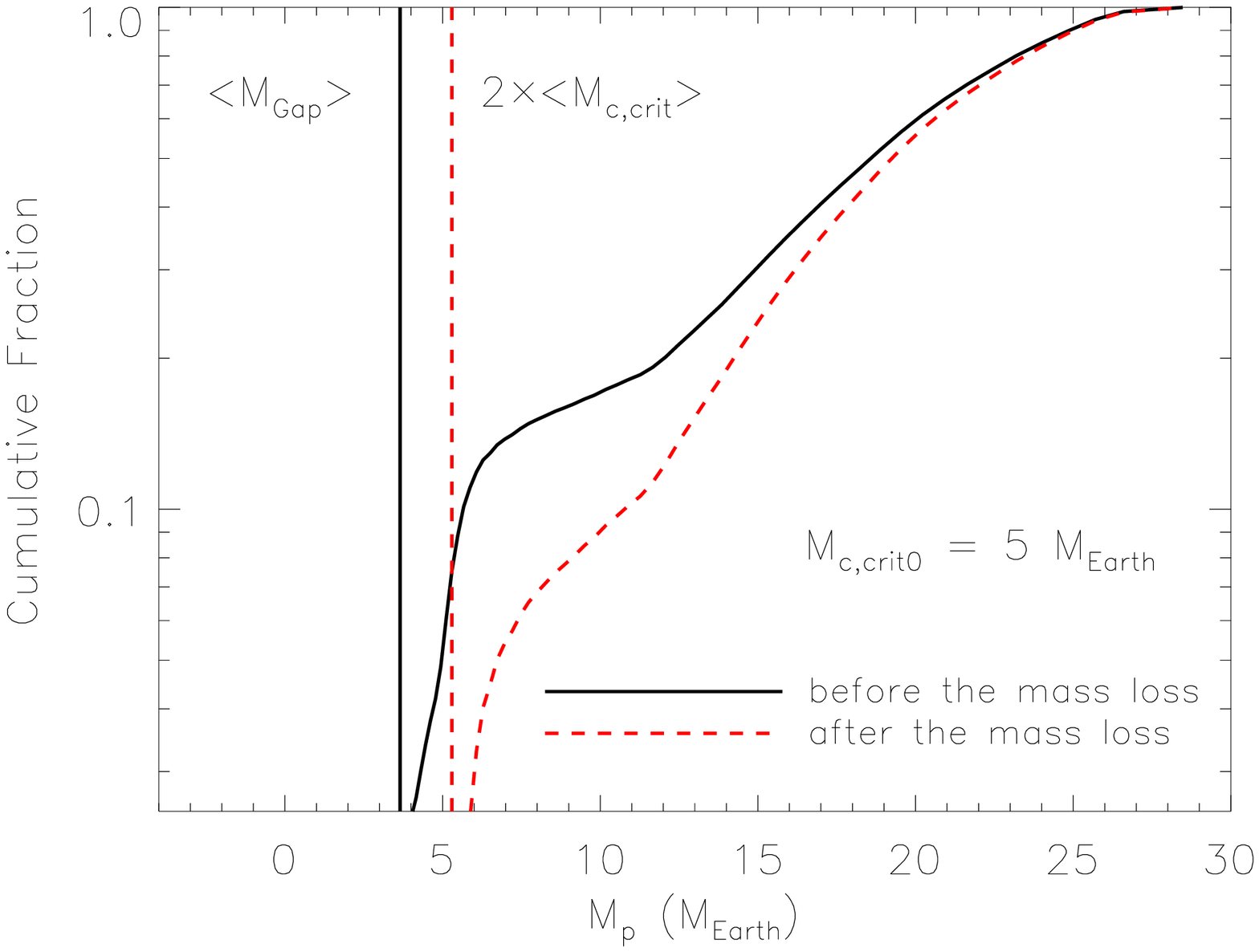}
\includegraphics[width=8cm]{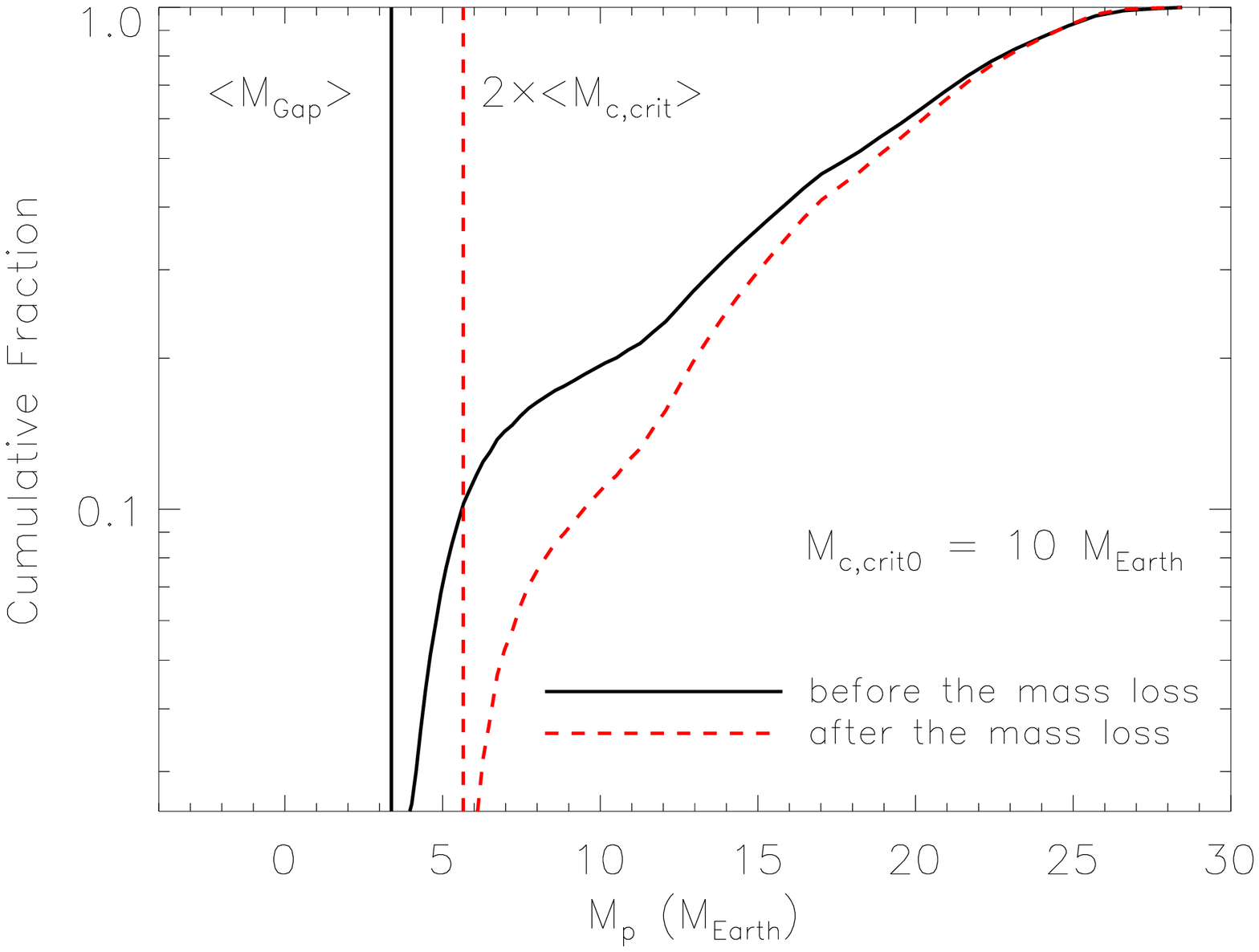}
\caption{The cumulative fraction of the normalized PFFs as a function of planetary mass.
The results for the case that $M_{c,crit0}=5 M_{\oplus}$ are shown on the left panel,
while those for the case that $M_{c,crit0}=10 M_{\oplus}$ are on the right one. 
The fraction before photoevaporative mass loss is denoted by the black, solid line, 
and that after the mass loss is by the red, dashed line.
Our results show that the fraction increases rapidly when $M_{p} \simeq 4-5 M_{\oplus}$ 
for the case of no photoevaporative mass loss (see the black, solid line). 
The mass is defined as $M_{min}^{CA}$ (see Table \ref{table5}).
We find that the value of $M_{min}^{CA}$ is insensitive to a planet formation parameter (i.e., $M_{c,crit0}$).
Instead, it is characterized well by the value of $\braket{M_{gap}}$.
This implies that switching of migration modes regulates the value of $M_{min}^{CA}$.
For the photoevaporative mass loss case, 
the fraction is computed only for planets that can keep some atmospheres even after the mass loss.
Our results demonstrate that the mass loss reduces the fraction at $5 M_{\oplus} \la M_{p} \la 20 M_{\oplus}$,
while the overall trend of the fraction is similar to the case of no photoevaporative mass loss (see the red, dashed line).
This reduction ends up with a higher value of the minimum mass of planets surviving the mass loss.
The mass is defined as $M_{min}^{CA+PE}$ (see Table \ref{table5}).
We obtain that $M_{min}^{CA+PE} \simeq 6 M_{\oplus}$ for both the cases of $M_{c,crit0}$.
This arises because $M_{min}^{CA+PE}$ is represented well by $\sim 2 \times \braket{M_{c,crit}}$.
}
\label{fig3}
\end{center}
\end{minipage}
\end{figure*}

\begin{table*}
\begin{minipage}{17cm}
\begin{center}
\caption{Characteristic masses of planets in the Low-mass regime}
\label{table5}
\begin{tabular}{clc}
\hline
Symbol                        &  Meaning                                                                                              & Values                                   \\ \hline
$M_{min}^{CA}$          &  Minimum mass of planets                                                                   & $\sim 4.5-5 M_{\oplus}$$^a$               \\
$M_{min}^{CA+PE}$   &  Minimum mass of planets surviving photoevaporative mass loss      & $\sim 6.6-6.9 M_{\oplus}$$^a$              \\
$M_{max}^{EA}$         &  Maximum mass of planets formed via the embryo assembly scenario              & $\sim 10 M_{\oplus}$              \\
\hline
\end{tabular}

$^a$ the value is estimated at the cumulative fraction of 0.05 (see Figure \ref{fig3}). \\
Note that this specific choice of the cumulative fraction does not affect the resultant value significantly, \\
since there is a sharp rise in the fraction for both the cases with and without photoevaporative mass loss (see Figure \ref{fig3}).
\end{center}
\end{minipage}
\end{table*}

To this end, we re-examine the PFFs of Low-mass planets (see Table \ref{table4}).
More specifically, we compute a cumulative fraction of the normalized PFFs as a function of planetary mass.
Figure \ref{fig3} shows the resultant behavior.
While the results for the case that $M_{c,crit0} = 5M_{\oplus}$ are presented on the left panel,
those for the case that $M_{c,crit0} = 10M_{\oplus}$ are on the right one.
We find that the cumulative fraction is an increasing function of planetary mass.
It is crucial that the fraction rises rapidly when the planetary mass exceeds $\simeq 4-5 M_{\oplus}$.
We hereafter refer to the value as "the minimum mass of planets",
which is labeled as $M_{min}^{CA}$ (see Table \ref{table5}).
We confirm that such a rapid increase occurs at a similar value of $M_{min}^{CA}$
for both the cases that $M_{c,crit0}=5 M_{\oplus}$ and $M_{c,crit0} =10 M_{\oplus}$.
Our results therefore indicate that there may be the minimum mass of planets that can be classified as the Low-mass planets,
when they are formed predominantly at planet traps.

We then investigate which physical process(es) can play an important role in determining the value of $M_{min}^{CA}$ for planets formed in our model.
First, one may wonder whether or not such a value may involve the critical core mass ($M_{c,crit}$ see equation (\ref{eq:m_ccrit})),
since the quantity regulates the onset of gas accretion (see Table \ref{table2}).
As already pointed out above, 
Figure \ref{fig3} shows that the threshold value of $M_{min}^{CA}$ is comparable even if $M_{c,crit0}$ varies.
Our parameterized approach of $M_{c,crit0}$ therefore leads to the consideration that 
the critical core mass and the subsequent gas accretion are not so central to quantify the value of $M_{min}^{CA}$ in our present calculations.
Second, we consider a possibility that the value of $M_{min}^{CA}$ may be relevant to planetary migration.
This is motivated by the results on the movement of planet traps (see Sections \ref{resu_1} and \ref{resu_2}).
To proceed, we compute the statistical average value of $M_{gap}$ (see equation (\ref{M_gap_ave})).
Figure \ref{fig3} shows the resultant value of $\braket{ M_{gap} }$ that is denoted by the vertical solid line there (also see Table \ref{table4}).
As expected, the sudden increase of the cumulative fraction is characterized well by the value of $\braket{ M_{gap} }$;
the majority of planets that end up in the Low-mass planet regime are more massive than $\braket{ M_{gap} }$.
In other words, type II migration was effective for them to be transported into their final orbital radius ($r < 1$ au).

Thus, we can conclude that, when planet traps play an important role in forming planets in gas disks,
switching of migration modes (from trapped to type II) can be used 
to estimate the value of $M_{min}^{CA}$ for planets that can fill in the Low-mass planet regime.

\subsection{Effect of photoevaporative mass loss} \label{resu_5}

We have so far focused on the consequence arising from the process of formation and migration of planets.
As mentioned in Section \ref{mass_loss}, it is interesting to investigate the fate of fully formed planets.
This is, in particular, important for close-in super-Earths,
partly because they will receive a significant amount of stellar photons due to their small orbital distance from the central star,
and partly because it is expected that they are more subject to such mass loss due to a low value of planetary masses. 
In fact, we have confirmed that most planets that experience photoevaporative mass loss are less massive 
and located at the distance of $r \la 0.1$ au in our model.
Also, most planets in the Low-mass regime have cores that are more massive than $1M_{\oplus}$ (see Table \ref{table4}).
Then, the inclusion of photoevaporative mass loss does not alter the value of the (integrated) PFFs significantly.
We therefore focus on a cumulative fraction of the normalized PFFs as done in the above section.

Figure \ref{fig3} shows the results of the cumulative fraction as a function of planetary mass (see the red, dashed line).
In the plot, we sum up only planets that experience photoevaporate mass loss, but keep some fraction of their atmosphere.
In other words, we exclude planets that completely lose their atmosphere.
We have adopted this criteria,
because then the results with and without photoevaporative mass loss can readily be compared.
Our results show that the general trend of the cumulative fraction does not change very much
even if photoevaporative mass loss is included in the model; the fraction increases steadily with planetary mass.
There are two main differences with the case of no photoevaporative mass loss (see the black, solid line).
First, the resultant value of the cumulative fraction becomes lower for the case of photoevaporative mass loss.
This occurs around $5 M_{\oplus} \la M_{p} \la 20 M_{\oplus}$,
and is a direct reflection of the mass loss.
More specifically, some planets in the mass range can undergo the partial or complete removal of their atmosphere.
This leads to the re-distribution of such planes from the mass bin where they have originally been located to the new, lower mass bin,
and eventually reduces the cumulative fraction there.

Second, a threshold value of planetary mass at which the cumulative fraction suddenly increases, 
shifts to a higher value.
This is another manifestation of the photoevaporative mass loss of planetary atmospheres.
As a result, the minimum mass of planets that contain some atmospheres becomes higher,
compared with the case of no photoevaporation.
We refer to such a threshold value as "the minimum mass of planets surviving photoevaporative mass loss",
which is labeled as $M_{min}^{CA+PE}$ (see Table \ref{table5}).
It is important that $M_{min}^{CA+PE} \simeq 6-7 M_{\oplus}$ 
for both the cases of $M_{c,crit0}=5 M_{\oplus}$ and $M_{c,crit0} =10 M_{\oplus}$.
This trend can be understood by two combined effects:
one of the effects involves a formation process;
the other is related to the feature of photoevporative mass loss.
For the former one, we find that 
planets that end up in the Low-mass regime have the value of $\braket{ M_{c,crit} } \la 3 M_{\oplus}$ 
for both the cases of $M_{c,crit0}$ (see Table \ref{table4}).
For the latter, the analytical mass loss model by \citet{lf13} is effective only for planets 
that satisfy the condition that $M_c \ga 0.5 M_p$ (see Section \ref{mass_loss}).
In the end, $M_{min}^{CA+PE}$ is characterized well by the value of $2 \times \braket{ M_{c,crit} }$ (see the vertical, dashed line).

Thus, our results indicate that, 
when planets form via the core accretion process coupled both with planet traps and with photoevaporative mass loss,
it is likely that planets with the mass of $\sim 6-7 M_{\oplus}$ or higher can keep some gaseous atmospheres.

\section{Discussion} \label{disc}

We have demonstrated above that planets in the Low-mass regime have some atmospheres 
when they are more massive than $M_{min}^{CA}$ which is $\sim 4-5 M_{\oplus}$.
Such a threshold value increases to $M_{min}^{CA+PE}$ of $\sim 6 -7M_{\oplus}$ 
when photoevaporative mass loss is taken into account (see Table \ref{table5}).
These results, however, are obtained under the assumption in which some physical processes are neglected and/or simplified.
We here discuss how such processes may affect our results.
The processes include planetary migration (Section \ref{disc_1}), planet-planet interactions (Section \ref{disc_2}), 
gas accretion and the subsequent mass loss (Section \ref{disc_3}), and dust physics occurring at ice lines (Section \ref{disc_4}).
We also briefly touch on other formation mechanisms of super-Earths (Section \ref{disc_5}).
Keeping these caveats in mind, 
we attempt to provide some implications of our results for observed close-in super-Earths (Section \ref{disc_6});
we couple our results with the mass-radius diagram of these planets.
It is interesting that the diagram can be viewed as an exoplanet "phase" diagram
in which some formation and evolution processes can be conjectured based on the mass and radius of observed planets.

\subsection{Planetary migration} \label{disc_1}

As demonstrated in Section \ref{resu_2}, planet traps can become effective in our model when $M_{mig,I} < M_p < M_{gap}$
even if the effect of (partial) saturation is taken into account.
And when $M_{gap} < M_p$, saturation can weaken corotation torque significantly,
so that Lindblad torque plays the major role in regulating planetary migration.
As a result, planets will undergo inward type II migration afterwords.

This assumption may be reasonable in our framework,
wherein all the physical processes are modeled as simple semi-analytical formulae.
It is nonetheless important to point out that more detailed simulations would be needed to fully confirm our results.
This is partly because the physics of saturation is not completely understood,
and partly because switching of migration modes and the subsequent type II migration are also far from a complete understanding.
For instance, some previous studies quantify the effect of saturation based on a simple timescale argument \citep[e.g.,][]{dmk14},
and others use the results of numerical simulations \citep[e.g.,][]{mc10,pbk10}
in which rather ideal situations are adopted to examine how saturation occurs.
In reality, planetary migration will occur simultaneously with planetary growth.
Recently,  \citet{bmk15} have numerically investigated the effect of planetary accretion on type I migration.
They have shown that, when the solid accretion rate onto planets is high enough,
the resultant accretion luminosity of planets can modify the density and thermal structure of disks especially in the vicinity of the planets.
They have found that such modifications are significant to reverse the direction of migration.
This newly discovered torque may be effective for planets of $\sim 0.5-3 M_{\oplus}$.
It is important that this range of planet mass corresponds roughly to the mass range at which saturation becomes effective.
Thus, it would be important to numerically investigate the effect of saturation in a more realistic situation.

For switching of migration modes and the subsequent type II migration, 
the presence of the disk gas in the horse-shoe orbit and the shape of partial gaps would play an important role.
As an example, \citet{mp03} claim that the residual gas in the horse-shoe orbit may excite the so-called type III migration.
This mode of migration is much faster than the rapid type I migration, and its direction is sensitive to the gas motion in the horse-shoe orbit.
Another example may be related to type II migration.
\citet{hi13} argue that type II migration is also fast enough to jeopardize the existence of gas giants beyond 1 au,
since its migration speed is determined by the local viscous timescale.
\citet{dk15} have undertaken the follow-up study, by performing numerical simulations,
and confirmed the rapidness of type II migration.
Moreover, their numerical simulations show that the behavior of type II migration depends sensitively on the structure of  (partial) gaps.
Thus, detailed simulations would be needed to realistically model how migration modes change from the type I to the type II regime,
and how type II migration proceeds in evolving disks.

In summary, while there are still a number of uncertainties in planetary migration,
our results may still be useful in the sense that the results may provide a conservative estimate on planetary populations.

\subsection{Planet-planet interactions} \label{disc_2}

One of the crucial simplifications in our model is that planet-planet interactions are not included.
It is obvious that our model is highly idealized on this aspect;
while formation of a planet occurs simultaneously at three traps in single disks,
our current model assumes that all the three planets are built independently.
In fact, the {\it Kepler} results suggest that many of super-Earths are multiple systems \citep[e.g.,][]{lff11,brb12}.
For these systems, $N-$body dynamics would play an important role in establishing the current architecture of orbital distributions. 

It may nonetheless be important to point out that 
planets formed in our model would be a "first" generation of close-in super-Earths,
since their formation takes place in gas disks.
This in turn suggests that, even if another processes such as giant impact can trigger the formation of super-Earths after gas disks are gone,
the main ruler of the systems would be the planets formed by the failed core scenario.
In addition, the formation and the evolution of close-in super-Earths would be 
affected considerably by the existence of gas giants in the systems \citep[e.g.,][]{min13}.
Our population synthesis calculations show that many gas giants can be present around $r= 1$ au (see Table \ref{table4}).
It is obviously important to fully couple our model with the computation of planetary dynamics.
We will investigate this effect in a subsequent paper (Hasegawa et al in prep).

\subsection{Gas accretion and the subsequent mass loss} \label{disc_3}

It would be interesting to investigate the process of gas accretion onto planetary cores in detail.
As an example, we here consider the effect of the heat content of planetary cores.
In this paper, we have assumed that the heat content of planetary cores is negligible 
when gas accretion onto the cores proceeds.
This assumption may be appropriate when planetary cores are massive enough 
to eventually trigger runaway gas accretion to form gas giants \citep[e.g.,][]{lc15}.
For close-in super-Earths, however, this assumption may not hold; 
\citet{ih12} show that low-mass cores of super-Earths can experience significant erosion of their envelope during disk dispersal.
This occurs because the heat content of such cores can expand their atmosphere and prevent the gas from being accreted onto the cores.
Furthermore, rapid dispersal of disk gas that can be caused by photoevaporation of disks
can serve as an additional agent to slow dow gas accretion.
This becomes possible because gas supplies from the disks to the cores can be reduced considerably.
Thus, it would be interesting to use a more sophisticated model for the gas accretion process, 
and to explore how the resultant populations of close-in super-Earths can be affected.

Also, it would be important to improve the analytical model of photoevaporative mass loss.
While the current model enables a seamless coupling with our population synthesis calculations due to its simplicity (see Section \ref{mass_loss}),
there are two drawbacks \citep{lf13}:
the one is that the model can work effectively only for planets that $M_c \ga 0.5 M_p $;
the other is that the model cannot trace the evolution of a tiny ($\la 1-10 \%$) mass fraction of envelopes surrounding planetary cores.
These limitations arise because the analytical model becomes less valid when non-linear effects become more important,
which is the case for the above two situations.
Since the typical envelope mass fraction of observed super-Earths is very likely $\sim 1-10 \%$ \citep[e.g.,][]{wl15},
it would be crucial to use a more sophisticated model for a more careful examination of these planets.

\subsection{Dust physics at ice lines} \label{disc_4}

We have so far adopted a model of \citet{hp11} to compute the position of ice line traps.
In the model, water-ice lines are specified under the assumption 
that viscous heating plays a major role in determining their distance from the central star.
Recently, \citet{pob15} have pointed out that the location of ice lines can move inward by some amount ($\sim$ up to 50 \%).
This inward movement is triggered both by the radial drift of dust particles and by disk accretion onto the central star.
While the movement of ice line traps that is driven by disk accretion is already included in our model (see equation (\ref{eq:r_il})),
it would be interesting to calibrate how the resultant planetary population will vary 
due to the additional inward movement triggered by the radial drift of dust particles.
Since the further inward movement of ice line traps decreases $M_{gap}$ (see Figure \ref{fig2}),
we expect that this effect may decrease the value of $M_{min}^{CA}$ somewhat.

\subsection{Other formation mechanisms of super-Earths} \label{disc_5}

Based on our results (see Table \ref{table5}), 
the failed core scenario coupled with planet traps would work well for planets that have more massive than 4-5 $M_{\oplus}(=M_{min}^{CA})$.
The resultant value of $M_{min}^{CA}$ is apparently consistent with the observational data (see Figure \ref{fig1}, also see Section \ref{disc_6}).
Note that the observed super-Earths that have the 2-$\sigma$ mass determination are only selected in Figure \ref{fig1}.
One may then wonder what happens if all the Kepler data are taken into account;
while the current observations can marginally detect the mass of super-Earths observed by the Kepler satellite \citep{wm14},
it seems that most of them have the radius of $\ga 1.6 R_{\oplus}$, but may have the mass of $\la 5 M_{\oplus}$.
As clearly mentioned in Section \ref{resu_3}, our model can generate only some fraction of observed super-Earths (see Table \ref{table4}).
This indicates that observed super-Earths should be formed by a number of mechanisms (see Section \ref{intro}).
For instance, planets formed via embryo assembly can become as massive as $\sim 1 M_{\oplus}$ 
and up to $\sim 10 M_{\oplus}$ \citep[e.g.,][]{oi09,hm13},
while its value depends on how much of solids can accumulate in the inner region of disks.
Also, \citet{lco14} have recently shown that,
if the gas disk lifetime is as short as $\sim1$ Myr, 
the typical envelope mass fraction of observed super-Earths can be
reproduced well by computing gas accretion onto protoplanets that form either via in situ or via embryo assembly \citep[also see][]{lc16}.
Thus, the failed core scenario coupled with planet traps may be applicable for super-Earths that are more massive than $4-5 M_{\oplus}$,
while other mechanisms may play a role for super-Earths that are less massive than the value (also see Section \ref{disc_6}).

\subsection{Implications for the mass-radius diagram} \label{disc_6}

Bearing the above caveats in mind, 
we discuss some implications that can be derived from our results.

\begin{figure*}
\begin{minipage}{17cm}
\begin{center}
\includegraphics[width=8cm]{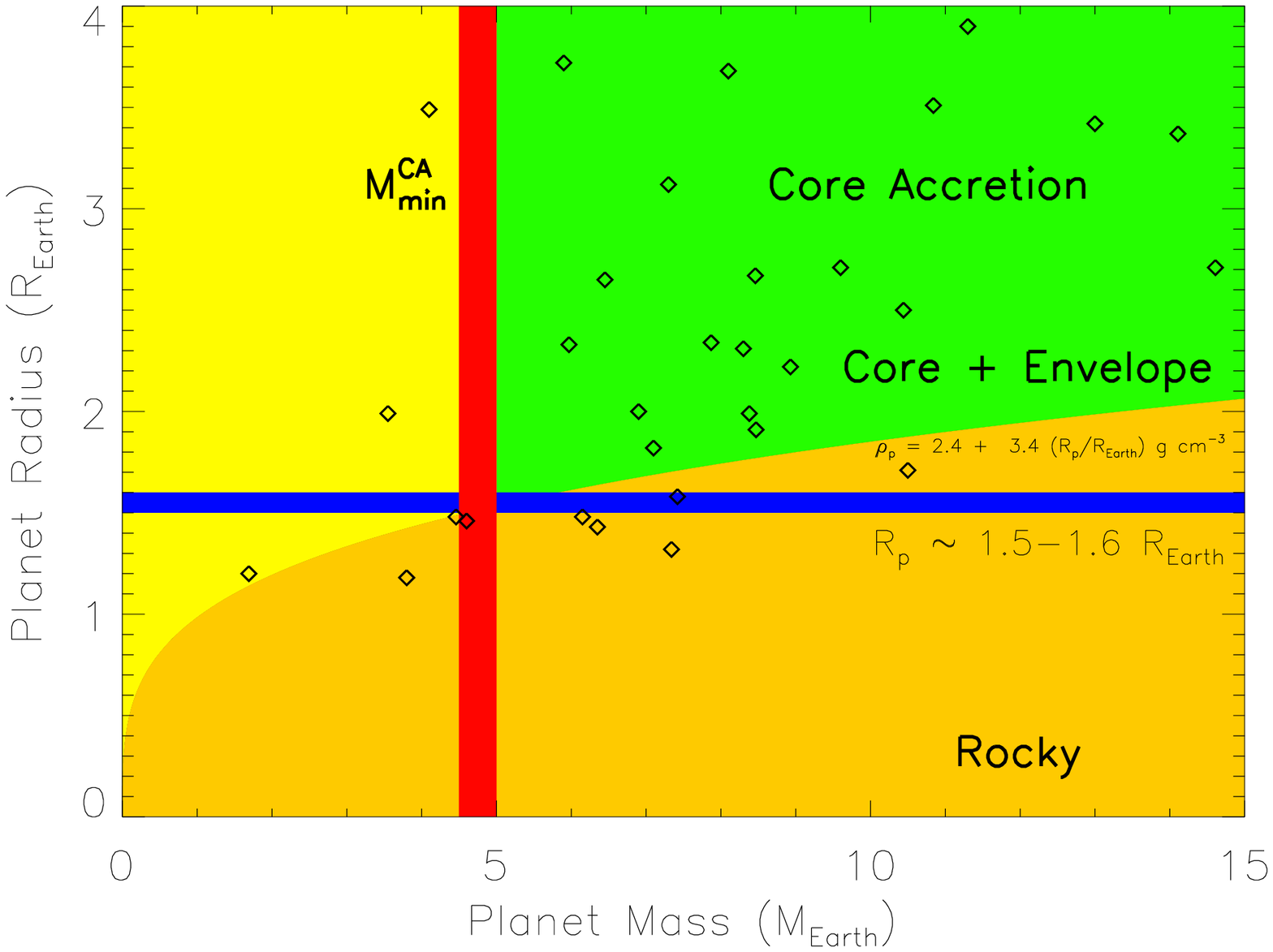}
\includegraphics[width=8cm]{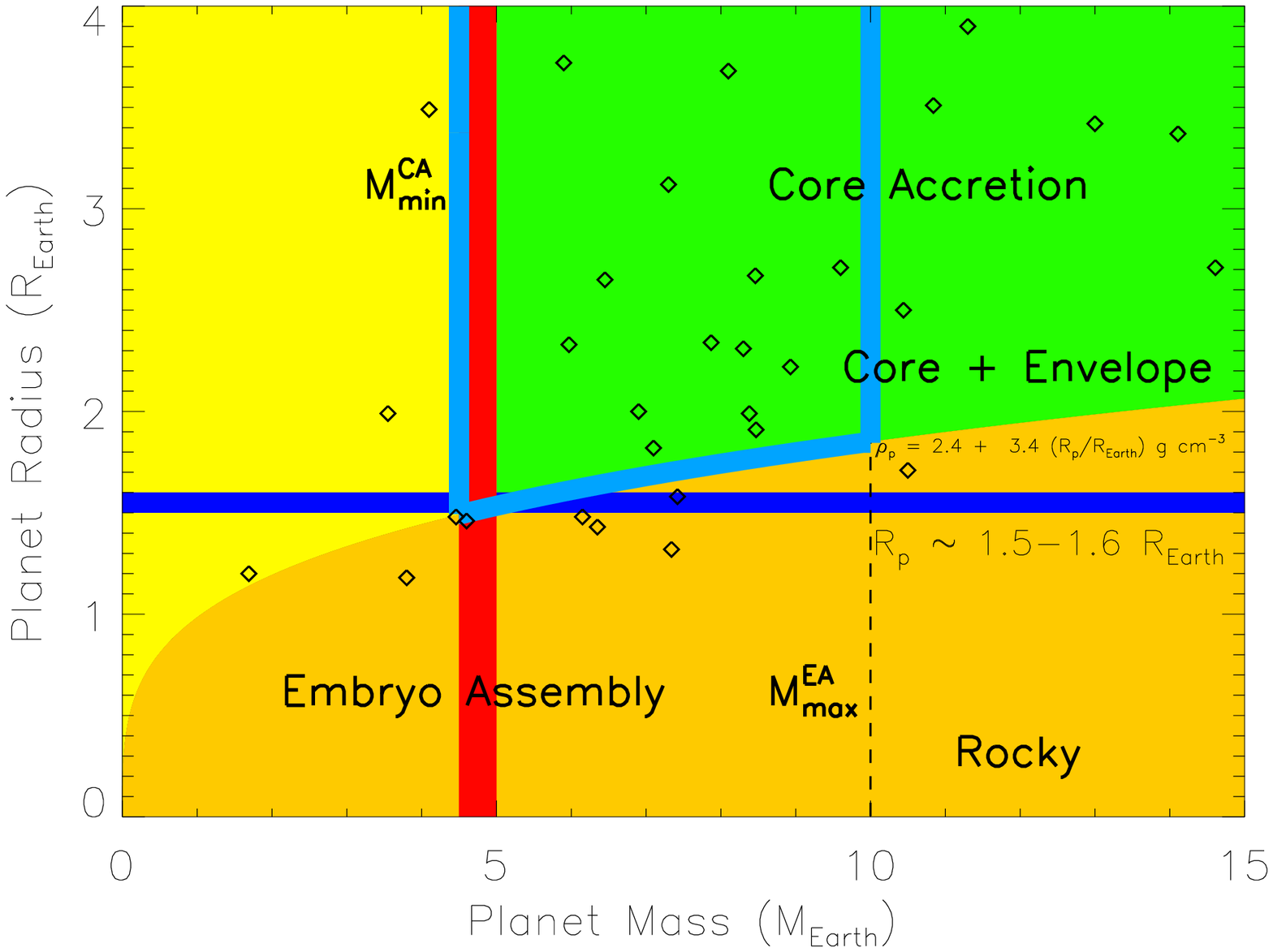}
\includegraphics[width=8cm]{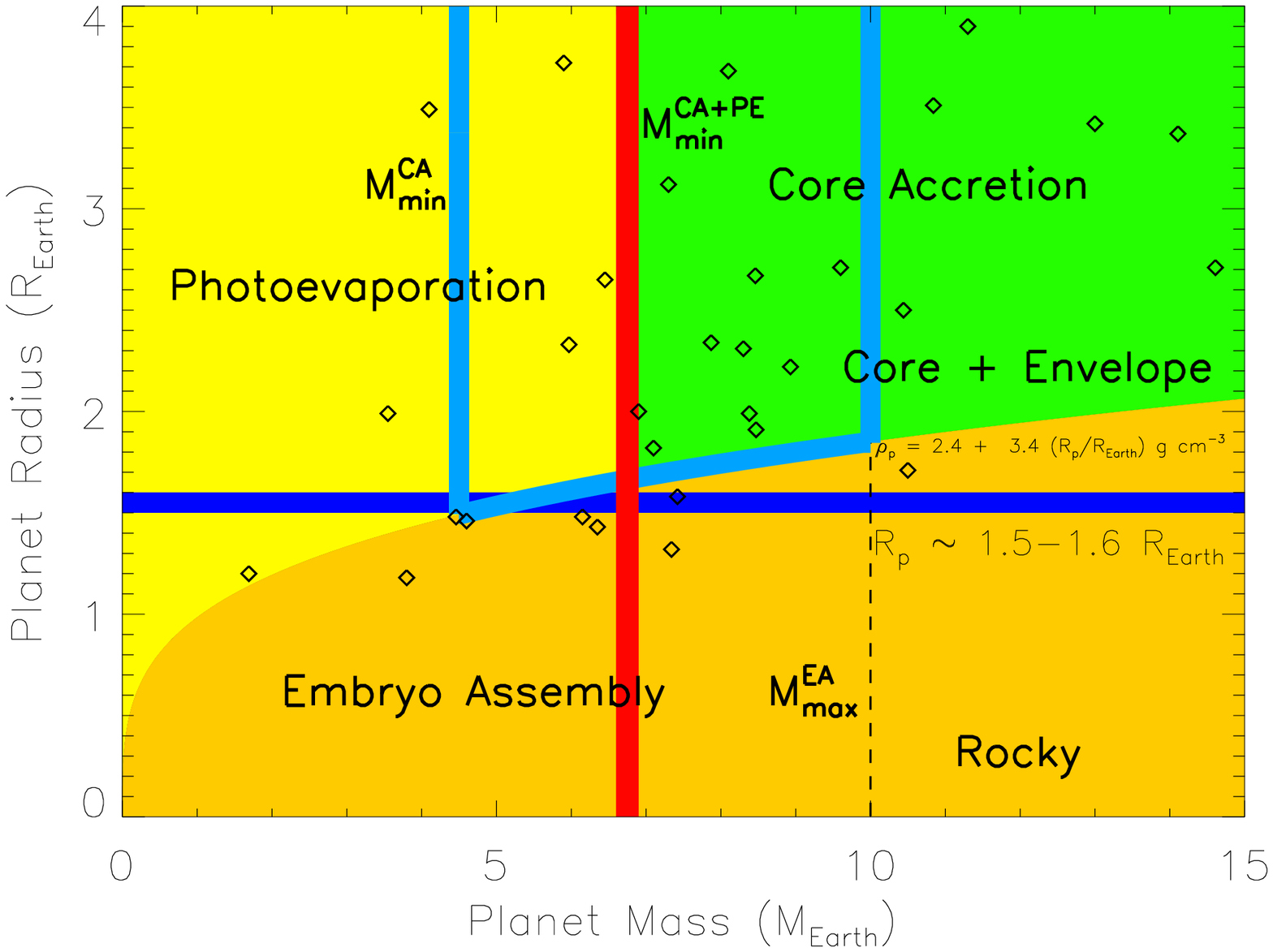}
\includegraphics[width=8cm]{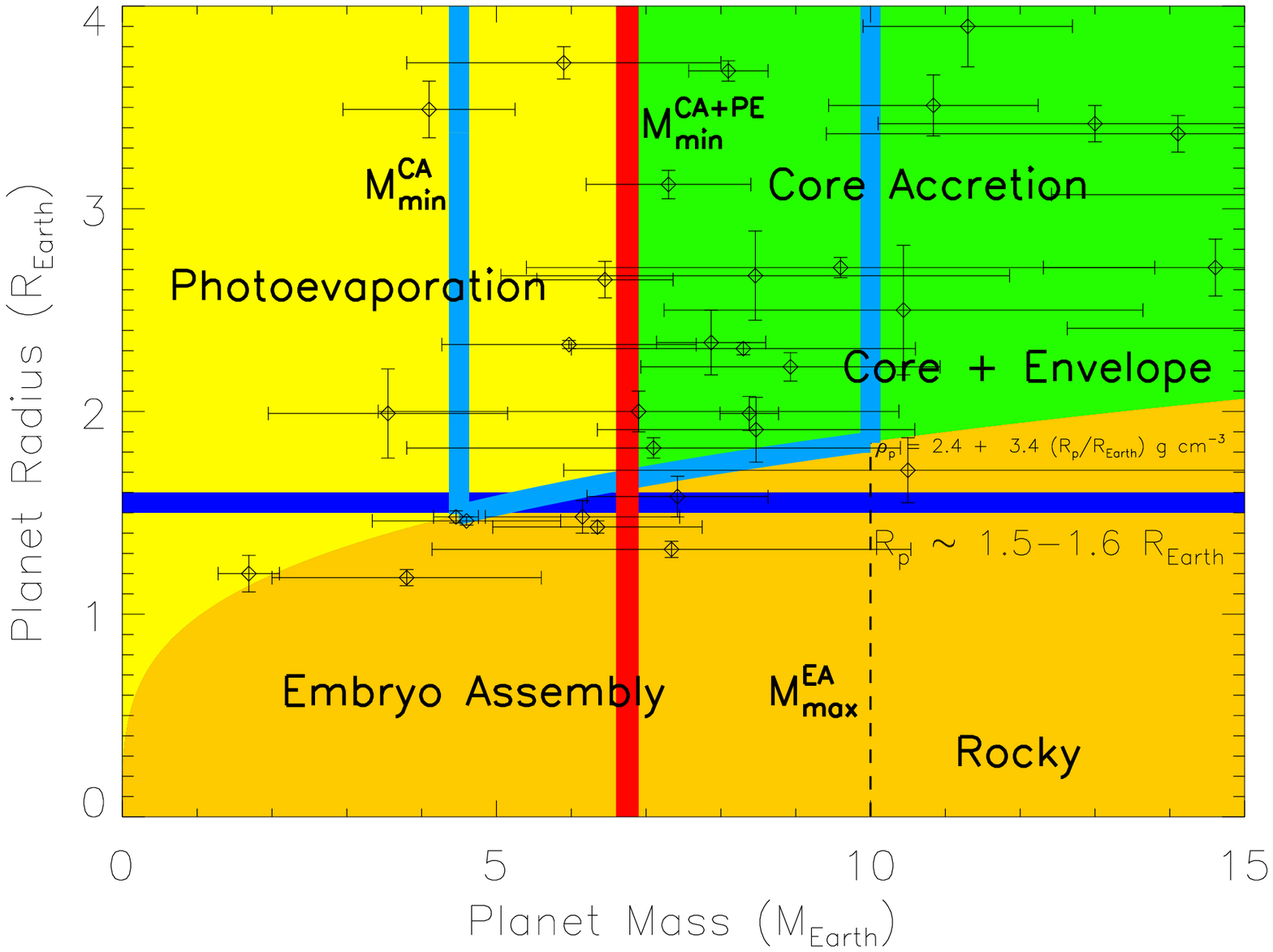}
\caption{Exoplanet "phase" diagram. On the upper left, the value of $M_{min}^{CA}$ is incorporated (see the vertical red line).
On the upper right, that of $M_{max}^{EA}$ is included (see the vertical dashed line).
On the lower left, that of $M_{min}^{CA+PE}$ is contained (see the vertical red line).
On the lower right, the error bars for observed exoplanets are explicitly labelled.
The observational data are adopted from \citet{mwp14} (as Figure \ref{fig1}).
These diagrams may be regarded as an exoplanet "phase" diagram,
since the formation and evolution modes can be speculated based on the mass and radius of observed exoplanets.
}
\label{fig4}
\end{center}
\end{minipage}
\end{figure*}

As discussed above, our model predicts that 
the minimum mass of planets ($M_{min}^{CA}$) is about $4-5 M_{\oplus}$ (see Table \ref{table5}).
This value is derived from our statistical analysis, 
and is a consequence of switching of migration modes from trapped type I to inward type II migration.
Since the failed core scenario provides significant chance for super-Earths to accrete the disk gas (see Section \ref{resu}),
it can be expected that planets that have masses larger than $M_{min}^{CA}$ should contain some gaseous envelopes.
Incorporating this outcome into the mass-radius diagram of observed close-in super-Earths (see Figure \ref{fig1}),
we can identify a parameter space 
in which the failed core scenario may play an important role in synthesizing planets there.
This is depicted in Figure \ref{fig4} (upper, left); 
all the planets beyond $M_{min}^{CA}$ should be made of solid cores with gaseous atmosphere (see the green regime).
It is interesting that the value of $M_{min}^{CA}$ (that is denoted by the vertical red line) corresponds roughly to the planet mass 
that is estimated at the transition radius \citep[$R_{tran} \simeq 1.5-1.6 R_{\oplus}$, also see the horizontal blue line, ][]{r15}.

Other formation mechanisms would be needed to fully reproduce the population of observed super-Earths 
(see Table \ref{table4}).
As pointed out in Sections \ref{resu_3} and \ref{disc_5},
one of the plausible mechanisms may be the embryo assembly scenario.
Assuming that the maximum mass of super-Earths formed by this scenario may be $\sim 10 M_{\oplus}$,
which is labeled as $M_{max}^{EA}$ (see Table \ref{table5}), 
we can add the vertical dashed line in Figure \ref{fig4} (upper, right).
Then we can discuss an intriguing implication: 
when planets are located beyond the light blue line in the green regime,
the failed core scenario is very likely to act as a dominant mechanism to form them.
In the yellow regime, the embryo assembly scenario may be important to synthesis planets there.
And in the green regime surrounded by the light blue line,
both the failed core and the embryo assembly scenarios may be effective to build planets.  
  
Finally, we can include the effect of photoevaporative mass loss from planets.
As demonstrated in Section \ref{resu_5},
inclusion of the mass loss ends up with $M_{min}^{CA+PE} \simeq 6- 7 M_{\oplus}$ (see Table \ref{table5}).
In other words, planets that are more massive than $M_{min}^{CA+PE}$ can maintain their gaseous envelope.
As a result, it may be possible to modify the mass-radius diagram as what follows (see the lower left panel in Figure \ref{fig4});
when planets are located in $1 M_{\oplus} \la M_{p} \la M_{min}^{CA}$,
the embryo assembly scenario may be most likely.
When planets are in the range of $M_{min}^{CA} \la M_{p} \la M_{min}^{CA+PE}$,
both the embryo assembly and the failed core scenarios can be effective.
Since our statistical results indicate that photoevaporative mass loss can remove the entire atmosphere of planets (see the yellow regime),
planets in the range of $M_{p} \la M_{min}^{CA+PE}$ 
may eventually line up along the boundary between the yellow and the orange regimes
when the planets are formed via the failed core scenario coupled with planet traps.\footnote{
Note that planets formed via the embryo assembly coupled with the subsequent gas accretion do not necessarily line up along the boundary
as long as the planets can keep their atmosphere.}
For the range of $M_{min}^{CA+PE} \la M_{p} \la M_{max}^{EA}$,
both the mechanisms again can work together to fill out the regime.
Note that in the regime, photoevaporative mass loss may not be so crucial for planets formed via the failed core scenario.
As planetary mass increases and planets are located in the range of $M_{p} \ga M_{max}^{EA}$,
then the failed core scenario may be most important to form them.

Thus, the coupling of our results with the mass-radius diagram can potentially serve as an exoplanet "phase" diagram.
This diagram may be useful to roughly specify the formation and evolution mechanisms of observed exoplanets 
before a more detailed modeling will be undertaken.
It is interesting that most of observed exoplanets with the 2-$\sigma$ mass detection are located within the error bars 
in the green regime as well as along the boundary between the yellow and orange regimes (see the lower right panel in Figure \ref{fig4}).

\section{Conclusions} \label{conc}

One of the key discoveries in exoplanet observations 
that are mainly done both by the radial velocity technique and by the transit method via the {\it Kepler} mission,
is the rapidly growing population of close-in super-Earths.
These samples are quite unique in the sense that their mass ($1 M_{\oplus} \la M_p \la 20 M_{\oplus}$) and semimajor axis ($r \la 1$ au) 
do not overlap with the parameter space covered by planets in the solar system (see Figure \ref{fig1}).
This therefore triggers active investigations of super-Earths 
in which a number of formation mechanisms have currently been examined (see Table \ref{table1}).

In this paper, we have considered a failed core scenario and examined planetary migration in detail.
In the scenario, the formation of super-Earths is regarded as a scaled-down version of gas giant formation.
As a result, the population of such planets should emerge at the late stage of disk evolution,
wherein core formation takes a longer time due to a lower value of the solid density. 
Such slow growth of planetary cores can eventually prevent the cores from undergoing rapid gas accretion.
In other words, the formation timing would become a central quantity to determine the feasibility of the failed core scenario.
Also, the failed core scenario can lead to the formation of massive cores in gas disks,
where rapid type I migration cannot be neglected.
This provides another difficulty in examining the scenario in detail,
because the direction of the migration is very sensitive to the local disk structures.
Here, we have focused on planet traps - the specific location of protoplanetary disks at which rapid type I migration can be halted.
We have investigated how plausible the failed core scenario is to account for the observed properties of close-in super-Earths.
We have adopted a model that has been developed in our earlier work \citep[see Section \ref{model}]{hp11,hp12,hp13a}.

We have first examined the movement of planet traps (see Figure \ref{fig2}).
We have shown that, while planet traps can spread out over a wide range of disk radii at the early stage of disk evolution,
they end up around $r \simeq 1$ au at the stage where gas disks are severely depleted.
This occurs, because their movement is coupled directly with disk evolution (see equations (\ref{eq:r_dz}), (\ref{eq:r_il}), and (\ref{eq:r_ht})).
In other words, their position can freeze out once disk evolution terminates due to photoevaporation of the disk gas.
We have found that the final orbital distance of planet traps is beyond $r \simeq 1$ au even if a disk mass parameter varies considerably.
Thus, our results indicate that when planet traps can play an important role in forming planets,
planets should drop-out from their host trap to distribute within $r \simeq 1$ au.
We have then computed a number of characteristic masses at planet traps.
These include a threshold mass ($M_{mig,I}$) for planets to migrate fast enough to follow the moving of planet traps (see equation (\ref{eq:M_typeI})),
a critical planet mass ($M_{sat}$) for saturation to become strong enough to null the effect of planet traps (see equation (\ref{eq:M_sat})), 
the gap-opening mass ($M_{gap}$) of planets (see equation (\ref{eq:M_gap})), 
and a planet mass ($M_{mig,II}$) that becomes comparable to the value of $2 \pi r^2 \Sigma_g$ (see equation (\ref{eq:M_typeII})).
We have found that planet traps become effective for planets with certain masses ($M_{mig,I}< M_p < M_{gap}$)
even if the effect of (partial) saturation is taken into account.
This is because $M_{gap} \approx M_{sat}$.
We have also confirmed that $M_{gap} < M_{mig,II}$, indicating that planets will undergo type II migration once their mass exceeds $M_{gap}$.
Moreover, we have demonstrated that the value of $M_{gap}$ at the smallest orbital distance of planet traps is about $2-3 M_{\oplus}$ (see Figure \ref{fig2}).
This implies that when planets can grow up to this threshold value or higher, 
they can drop-out from their planet traps and be transported to the vicinity of their central star via inward type II migration.
We have confirmed that the value of $M_{gap}$ is not so sensitive to the disk mass parameter ($\eta_{acc}$).

We have also performed a population synthesis analysis 
to investigate how useful the failed core scenario is to understand the properties of observed close-in super-Earths.
We have initially examined the resultant PFFs of Jovian planets,
and demonstrated that the coupling of planet traps with the core accretion scenario can lead to planetary populations,
which are in a good agreement with the trend of exoplanet observations (see Table \ref{table4});
the PFFs of exo-Jupiter are much higher than those of hot Jupiters.
On the other hand, we have found that the PFFs of the Low-mass planets are not fully consistent with the observations 
which reveal that close-in super-Earths are currently the most dominant.
This difference probably suggests that a number of formation mechanisms would be needed 
to reproduce the entire population of observed super-Earths.
We have then turned our attention to the population of the Low-mass planet regime (see Table \ref{table3}),
and examined how these planets are formed in our model.
To proceed, we have plotted a cumulative fraction of the normalized PFFs as a function of planetary mass (see Figure \ref{fig3}).
We have found that the fraction rises rapidly 
when the planetary mass exceeds $4-5 M_{\oplus}$ (that is defined as $M_{min}^{CA}$) (see Table \ref{table5}).
As discussed above, the value is characterized well by the statistical average value of $M_{gap}$ ($\braket{M_{gap}}$).
Our results therefore indicate that switching of migration modes (from trapped type I to type II migration)
is needed for planets that grow predominantly at planet traps to distribute in the vicinity of the central star.
It is important that the value of $M_{min}^{CA}$ is independent of planet formation parameters such as the critical core mass ($M_{c,crit}$).
Thus, we can conclude that planets with the mass of $M_{min}^{CA}$ or higher 
can be made of solid cores surrounded by gaseous atmospheres (see Figure \ref{fig4}).

In addition, we have examined the effect of photoevaporative mass loss from planets (see equation (\ref{eq:f_lost})).
We have found that the mass loss decreases the cumulative fraction of planets,
which corresponds to the range of planetary mass ($1 M_{\oplus} \la M_p \la 20 M_{\oplus}$) (see Figure \ref{fig3}).
This is a direct reflection that planets in the range undergo the partial or the complete removal of their envelope.
As a result, a threshold mass of planets at which the fraction suddenly increases, 
goes up to $6-7 M_{\oplus}$ (that is defined by $M_{min}^{CA+PE}$) (see Table \ref{table5}).
Thus, our results imply that planets with the mass of $M_{min}^{CA+PE}$ or lower may not currently have gaseous atmosphere 
due to photoevaporative mass loss,
when the planets are formed via the failed core scenario coupled with planet traps.

We have discussed physical processes that are simplified and/or not included in our present model (Section \ref{disc}).
These include planetary migration, planet-planet interactions, 
gas accretion and the subsequent mass loss, and dust physics at ice lines.
While some of them may have a considerable impact on our results,
our findings may still be valuable in the sense that they may provide a conservative estimate on the resultant planetary populations.
Finally, we have provided some implications that can be obtained from our results (see Figure \ref{fig4}).
Since our analysis may be useful to derive the threshold value of planetary mass (see Table \ref{table5}),
the coupling of our results with the mass-radius diagram may act as an exoplanet "phase" diagram
in which some formation and evolution processes can be speculated based on the "location" of planets in the diagram.

Overall, the failed core scenario coupled with planet traps can reproduce some fraction of observed close-in super-Earths,
and may be important in the sense that it can serve as one of the crucial mechanisms 
to fully understand the diversity of observed properties of such planets.

In a subsequent paper, planet-planet interactions should be incorporated into our model,
in order to examine how our current results can be affected.


\acknowledgments
The author thanks Matthew Alessi, Ramon Brasser, Pin-Gao Gu, Soko Matsumura, Neal Turner, and Ralph Pudritz for stimulating discussions,
and two anonymous referees for useful comments on my manuscript.
The author is currently supported by Jet Propulsion Laboratory, California Institute of Technology,
and has previously been by EACOA Fellowship that is supported by East Asia Core Observatories Association which consists of 
the Academia Sinica Institute of Astronomy and Astrophysics, the National Astronomical Observatory of Japan, the National Astronomical 
Observatory of China, and the Korea Astronomy and Space Science Institute.  
The part of this research was carried out at JPL/Caltech, under a contract with NASA.






\bibliographystyle{apj}          

\bibliography{apj-jour,adsbibliography}    

\end{document}